\documentstyle[aas2pp4]{article}

\def\sb{\ifmmode{\;{\rm mag}\;{\rm arcsec}^{-2}}\else{~mag~arcsec$^{-2}$}\fi}
\def\csb{\ifmmode{\mu_0}\else{$\mu_0$}\fi}
\def\eg{e.g.,\ }
\def\ie{i.e.,\ }
\def\cf{cf.\ }
\def\vs{vs.\ }
\def\etal{{\it et al}.\ }

\begin{document}

\title{Gas Mass Fractions and the Evolution of Spiral Galaxies}
\author{Stacy S. McGaugh}
\affil{Department of Terrestrial Magnetism, Carnegie Institution of Washington}
\authoraddr{5241 Broad Branch Road, NW, Washington, DC 20015}
\and
\author{W. J. G. de Blok}
\affil{Kapteyn Astronomical Institute, University of Groningen}
\authoraddr{Postbus 800, 9700 AV Groningen, The Netherlands}

\lefthead{McGaugh \& de Blok}
\righthead{Spiral Galaxy Evolution}

\begin{abstract}

We show that the gas mass fraction of spiral galaxies is strongly
correlated with luminosity and surface brightness.
It is not correlated with linear size.
Gas fraction varies with luminosity and surface brightness at the
same rate, indicating evolution at fixed size.

Dim galaxies are clearly less evolved than bright ones,
having consumed only $\sim 1/2$ of their gas.
This resolves the gas consumption paradox, since there exist many galaxies
with large gas reservoirs.  These gas rich galaxies must have formed the
bulk of their stellar populations in the last half of a Hubble time.
The existence of such immature galaxies at $z = 0$
indicates that either galaxy formation is a lengthy or even ongoing process,
or the onset of significant star formation can be delayed for
arbitrary periods in tenuous gas disks.

\end{abstract}


\vspace{2in}
\centerline{Accepted for Publication in the {\it Astrophysical Journal}}


\section{Introduction}

The evolutionary history of galaxies can be inferred in a variety of ways.
One approach is to utilize the extensive knowledge of the
evolution of stars gleaned from the HR diagram to construct model galaxies
(\eg Tinsley\markcite{T68} 1968; Bruzual \& Charlot\markcite{BC} 1993).
More recently,
it has become possible to infer related information in a statistical
way from deep redshift surveys (\eg Lilly\markcite{CFRS} \etal 1995;
Ellis\markcite{ECBHG} \etal 1996).
Piecing the various lines of evidence together is not trivial, and rather
different pictures can emerge from equally valid approaches.

It is essential that the question be well posed.  What is meant by galaxy
evolution? We generally imagine that galaxies form out of the gas that emerges
from the hot big bang.  Regardless of the details of galaxy formation,
the units which we observe as galaxies today have converted some fraction of
this gas into stars.  The gas mass fraction $f_g$
is thus a natural parameter for quantifying the degree of evolution of a galaxy.

This obvious path has not been much pursued for the simple reason that most
well observed galaxies have low gas fractions ($f_g \lesssim 0.2$).
Their evolution is well advanced, and the small
dynamic range in $f_g$ provides no useful constraint on models.
Indeed, this leads to a conundrum in that we appear to live at a special time
when most disks are near to exhausting their gas reservoirs
(\eg Roberts\markcite{RX} 1963; Kennicutt\markcite{KTC} \etal 1994).

However, those objects which are well studied need not be representative.
It has recently been demonstrated that disk galaxies of low surface brightness
are quite common (McGaugh\markcite{MBS} \etal 1995a, de Jong\markcite{dJ3} 1996,
McGaugh\markcite{me} 1996a; Impey\markcite{imp} \etal 1996).
These galaxies are generally gas rich
(Schombert\markcite{lsbcat} \etal 1992; de Blok\markcite{dB2} \etal 1996),
and $M_{\rm HI}/L$ increases systematically with decreasing surface brightness
(Fig. 1).  This ratio depends only on the relative amounts of
atomic hydrogen and star light, and must be an evolutionary effect.
Attempting to understand this evolution and its strong dependence on surface
brightness is the subject of this paper.

\placefigure{mhi}

We emphasize that our discussion is restricted to disk galaxies, primarily
spirals.  We do not address Elliptical or dwarf Spheroidal galaxies,
though we do thoroughly examine relevant selection effects.
When required, distance dependent quantities assume
$H_0 = 75~{\rm km}\,{\rm s}^{-1}\,{\rm Mpc}^{-1}$.

\section{The Data}

The data consist of optical and 21 cm fluxes.
Data are selected to provide as large as possible a database for
investigating the relation between the gas content and optical properties
of spiral galaxies, particularly surface brightness.  A prerequisite is
that each of the quantities of interest are actually measured.  Surface
brightness, by which we mean the central surface brightness of the disk
determined from extrapolation to $R = 0$ of exponential fits to the
observed light profile, is not commonly
measured and is therefore the most limiting factor.  Of studies which
specifically measure surface brightness in the appropriate fashion,
HI measurements are not always available, further limiting the available data.
It would be desirable to also have CO measurements as a tracer of the
molecular gas, but there are essentially no published data which meet this
additional criterion, and low surface brightness galaxies have yet to
be detected in CO at all (Schombert\markcite{CO} \etal 1990).  Instead,
we will use the gas phase-morphological type relation of Young \&
Knezek\markcite{H2H1} (1989) molecular gas content.  A further piece
of information required for estimating the stellar mass is either an
$I$ magnitude or a $B-V$ color (see \S 3.1).  In sum, we need to know
$M_B$, \csb, $h$, $M_{\rm HI}$, the morphological type ${\cal T}$, and $B-V$ or $I$.

The large data sets which meet these requirements are those of
Romanishin \etal (1982\markcite{rom1};
1983\markcite{rom2}), McGaugh \& Bothun\markcite{MB} (1994),
de Jong \& van der Kruit\markcite{dJ1} (1994), de Jong (1995\markcite{dJt};
1996\markcite{dJ3}), and de Blok \etal (1995\markcite{dB1}; 1996\markcite{dB2}).
Surface photometry has been performed on all galaxy images to determine
the disk parameters \csb\ (central surface brightness) and $h$ (scale length),
as described in the various sources.
Most of the data for higher surface brightnesses ($\csb < 23~B\sb$) are from
de Jong, whereas most of that for $\csb > 23$ are from our own work.
An obvious gap visible in Fig.~1(b) at intermediate surface brightnesses
between these two data sets is filled in Fig.~1(a) by the data of Romanishin.
Where used, data for the special case of Malin 1 are taken from
Bothun\markcite{malin1} \etal (1987) and Impey \& Bothun\markcite{IB} (1989).

No trends with $M_{\rm HI}/L$ are obvious in any one data set alone. Only
upon combining them does the strong relation in Fig.~1 become apparent.
Data over this large dynamic range (a factor of 100)
in surface brightness have not previously been available.

\subsection{Selection Effects}

The data have been selected to be as inclusive as possible given the
demands placed on information content.  We make no claim that these
data are complete in any volume limited sense.  For our present purposes,
this is less important than having data representative of all morphological
types, luminosities, etc.  This is because we are more interested in the
evolution of galaxies as physical systems than in the number density of
any given type.  Nonetheless, the potential effects of selection upon
interpretation warrant further discussion.

The optical properties of the galaxies are plotted in Figs.~2 and 3.
These illustrate some important basic facts about spiral galaxies.
Disks exist over a large range in luminosity, surface brightness,
and size.  No particular value is characteristic for any of these;
each has some distribution (McGaugh\markcite{me}
1996a; de Jong\markcite{dJ3} 1996).  Lower surface brightness disks
tend to be bluer (Fig.~3), not redder as expected if they have fading
evolutionary histories (McGaugh \& Bothun\markcite{MB} \etal 1994).
The lack of red low surface brightness galaxies may itself be a
selection effect (McGaugh\markcite{me} \etal 1996a), but one still
wishes to understand why the ones we do know about are so very blue.

\placefigure{MSBH}

Flux limited surveys are most efficient at identifying the brightest ($L^*$)
and highest contrast ($\csb^*$) galaxies that exist in the intrinsic
distributions in a given passband.
As a direct consequence, galaxy catalogs are numerically
dominated by such objects far out of proportion to their actual numbers.
Samples selected for detailed study from such catalogs naturally tend to
concentrate on these most prominent galaxies.  We have avoided this by
augmenting such data with our own data for low surface brightness galaxies.
As a result, the combined data provide as comprehensive a view of spiral
galaxies in the field as is currently possible, spanning the full range
of disk morphologies from Sa to Sm/Im with $\csb < 25$ and $h > 1$~kpc.

\placefigure{SBVI}

Though this procedure adds sorely needed weight to late types and low
surface brightnesses, by no means does it guarantee a homogeneous
sample.  One indication of this is present in Fig.~2 as an apparent lack
of low luminosity, high surface brightness galaxies with $M_B > -18$ and
$\csb < 23$.  This gap is not apparent when \csb\ is plotted against $h$
(McGaugh\markcite{IAU} 1996b), and does not necessarily indicate a real
lack of such galaxies.  In effect, the sample of de Jong contains only
bright ($M_B < -18$) galaxies while our own contains only ones of low
surface brightness ($\csb > 23$).  The excluded region could well be
populated, but no data which meet all our information
requirements exist in this region of parameter space.

To test whether this can have a serious impact on our results, we perform
the analysis described below for both the entire sample, and a subsample
limited to $M_B < -18$.  Though the lever arm for fainter galaxies is of
course weaker, no serious differences occur among gas related parameters.
The reason for this is simple: the galaxies are optically selected.
Conclusions about the relation between optical parameters relevant to
selection ($M_B$, \csb, and $h$) are difficult to draw in the absence of
a rigorously complete sample.  However, the investigation into the evolution
of galaxies is not much hindered because a wide range of these parameters
are at least represented.  Moreover, the initial selection
of the galaxies is independent of $M_{HI}$.  These data tell us quite a
bit about the evolution of the galaxy types we do know about, and delimit
what may still be missing.

\section{Gas Mass Fractions}

The baryonic gas mass fraction is
\begin{equation}
f_g = {M_g \over {M_g + M_*}}
\end{equation}
where $M_g$ is the total mass in the form of gas, and $M_*$ that in stars.
To relate these to the observations, we need $M_* = \Upsilon_* L$ and
$M_g = \eta M_{\rm HI}$.  The parameter $\Upsilon_*$ is the mass to light ratio
of the stellar population, while $\eta$ plays an analogous role for the gas.
It accounts both for helium and any mass in metals and for gas phases other
than atomic.

With these definitions, it is straightforward to obtain
\begin{equation}
f_g = \left(1+ {\Upsilon_* L \over \eta M_{\rm HI}}\right)^{-1}.
\end{equation}
The flux ratio $M_{\rm HI}/L$ is distance independent
and has small errors in most cases.  However,
the conversion factors $\Upsilon_*$ and $\eta$ must be estimated
from other information.

\subsection{Stellar Population Mass to Light Ratios}

We wish to use the observable star light as an indicator of the stellar
mass.  Our stellar mass to light ratio $\Upsilon_*$ includes all mass
in stars living or dead since we
wish to distinguish between the gaseous and stellar mass reservoirs.
The value of $\Upsilon_*$ can be estimated from population synthesis
models, but these are highly uncertain and depend on many factors
like the IMF, star formation history, metallicity, gas recycling, the poorly
known behavior of some late stages of stellar evolution, etc.  Even
worse for our purposes, it depends iteratively on the quantity of
interest, $f_g$.  Though such models are still very useful, it is of
obvious interest to have independent means of constraining $\Upsilon_*$.

We use the dynamical constraints imposed by observations of the
vertical stellar velocity dispersion (Bottema\markcite{Bott1} 1993), as
generalized by de Blok \& McGaugh\markcite{dBM2} (1997).  This gives
\begin{equation}
\Upsilon_* = 1.936 \times 10^{0.4(B-V)} - 1.943
\end{equation}
in the $B$-band.  This expression is derived from the disk Tully-Fisher
relation with normalization imposed by the velocity dispersion observations.
There is some uncertainty in the absolute normalization, but
considerably less than in estimates lacking the boundary condition imposed by
dynamical measurements.  The color term has a residual dependence on population
synthesis which goes in the expected sense that redder galaxies have
higher $\Upsilon_*$.  This reproduces the observed trend well (Fig.~4)
even though one does not expect $B-V$ to be a perfect indicator of the
variation of $\Upsilon_*$ with the many underlying variables.

\placefigure{Bott}

The most important aspect of equation~(3) is the normalization
to dynamical constraints which give the most reliable estimate of the mass
contained in stars.  Without this, it is always possible to contrive models
in which the stars contain all or none of the baryonic mass, and $f_g$ remains
highly uncertain.  With it, $f_g$ is reasonably well determined.

The numbers returned by equation~(3) are consistent both with
stellar population models (\eg van den Hoek\markcite{bobby}
\etal 1996) and rotation curve analyses (Sanders\markcite{Bob} 1996;
de Blok \& McGaugh\markcite{dBM2} 1997) which are independent of the
velocity dispersion measurements.  As a further independent check of
our results, we will also compute $f_g$ from a purely population synthesis
estimate of $\Upsilon_*^I$ in the $I$-band.  The $I$-band is thought to
be the most robust indicator of stellar mass (Worthey\markcite{Guy} 1994),
though one still does not expect a universal value for all populations.
The models of Bruzual \& Charlot\markcite{BC} (1993) for a galaxy with
constant star formation suggest a value of $\Upsilon_*^I \approx 1.2$
over a wide range of ages, with fairly modest variation.  Though both
the absolute normalization and the universal applicability of this value
is uncertain, the use of this fixed number in the $I$-band gives
indistinguishable results from those of equation~(3) in the $B$-band
(Fig.~5).

\placefigure{Mstar}

\subsection{Gas Phases}

For the gas, $\eta$ must be corrected for both the hydrogen mass fraction $X$,
and the phases of gas other than atomic.  We assume a solar hydrogen mass
fraction, giving $\eta = X_{\sun}^{-1} = 1.4$.  Though not universally correct,
most of this is due to primordial helium.  Variations in helium and metal
content result in deviations from this value of less than 10\%.
This is a very small effect compared to that of other gas
phases.  Ionized gas in \ion{H}{2} regions and hotter plasma is of negligible
mass in spiral galaxies, but not so molecular gas.  Hence,
\begin{equation}
\eta = 1.4 \left[1+\frac{M({\rm H_2})}{M({\rm HI})}\right].
\end{equation}
The ratio of molecular to atomic gas varies systematically with Hubble type
(Young \& Knezek\markcite{H2H1} 1989).  We parameterize their mean relation by
\begin{equation}
\frac{M({\rm H_2})}{M({\rm HI})} = 3.7-0.8 {\cal T} + 0.043 {\cal T}^2
\end{equation}
(Fig.~6).  For very late types, we have forced the fit to approach
$M({\rm H_2})/M({\rm HI}) \rightarrow 0$ as ${\cal T} \rightarrow 10$.
Equation~(5) gives a good fit to the trend of the data of
Young \& Knezek\markcite{H2H1} (1989) except perhaps for very early types,
but there are few galaxies in the sample with ${\cal T} = 1$.
Of greater concern is the large amount of real scatter at each type ---
${\cal T}$ is by no means a perfect indicator of $M({\rm H_2})/M({\rm HI})$,
so equation~(5) is only a statistical estimator of the relative molecular
gas content.

The normalization of equation~(5) depends on the CO to H$_2$ conversion factor,
the precise value and universality of which is a matter of much debate.
The conversion factor appears to be metallicity dependent
(Wilson\markcite{Wils} 1995), and later types are generally more metal poor
than early types.  If one were to correct the noisy correlation of
$M({\rm H_2})/M({\rm HI})$ with ${\cal T}$ for the noisy correlation of
$Z$ with ${\cal T}$, it would have only a modest effect on equation~(5)
since the inferred variation in the conversion factor with metallicity
is fairly small compared to the range in Fig.~6.  Little will change because
early types are approximately solar metallicity and will receive no correction
from the canonical value of the conversion factor.  The latest types
are roughly $0.1\;Z_{\sun}$ (McGaugh\markcite{OHme} 1994a),
corresponding to a factor of 5 increase in
the conversion factor.  However, $M({\rm H_2})/M({\rm HI})$ is inferred
to be very small in late types, and a five fold increase in a small number
still results in a very small correction to the final $f_g$.
Rather than convolve together all these uncertainties, we simply use
equation~(5), realizing that this is probably the greatest uncertainty
in the process.

\placefigure{molec}

The uncertainty in the correction for molecular gas will have little impact
on our final results.  The correction is only important in early types, which
are relatively gas poor $f_g \lesssim 0.3$.  It is the late types which are
inferred to be gas rich, on the basis of the HI alone.  If we are missing
enormous amounts of molecular gas in late types
it will only make yet more gas rich galaxies which are
already inferred to be very gas rich.  Hence our qualitative results are
very robust, even if the quantitative details are subject to improvement
in the CO data.  Moreover,
it is only the ratio $\Upsilon_*/\eta$ that matters to our analysis
(equation~2).  This ratio does not vary much, and is not correlated with
surface brightness.  In order to affect our results, this ratio would have
to vary systematically with surface brightness by an order of magnitude
in just the right way to offset the strong observed trend in $M_{HI}/L$.
Not only does this seem unlikely, it simply can not happen.  The gas fractions
of low surface brightness galaxies can not be lowered by large $\Upsilon_*$
as these are not allowed by the rotation curve data (de Blok \&
McGaugh\markcite{dBM} 1996), nor can $\eta$ be raised for high surface
brightness galaxies where CO is readily detected.  Hence the large observed
variation in $M_{HI}/L$ must correspond to one in $f_g$.

\paragraph{The role of molecular and atomic gas:}
It is generally thought that star formation occurs only in molecular gas.
However, it appears that there is very little and perhaps no molecular gas in
the lowest surface brightness galaxies (Schombert\markcite{CO} \etal 1990).
We should therefore consider the possibility that stars form directly in
from atomic gas.  It is not obvious that this is necessary, though.
The star formation rates are low, and a very modest amount of
molecular gas might exist as an intermediary step.  Indeed, one could
argue that the lack of molecular gas inhibits star formation and keeps
the surface brightness low.  While these are interesting possibilities, they
do not matter to our present analysis, which is concerned only
with the global supply of gas and not the details of its local consumption.
In this context, it is perhaps best to think of the HI as a reservoir for
future star formation, regardless of whether molecular gas is a mandatory
intermediate step.  Indeed, gradual radial infall from the extended HI disks
of spirals may be the source of the inferred accretion of unprocessed material
since there is no evidence for substantial HI reservoirs in spherical halos
(Bothun\markcite{B85} 1985).

\section{The Relations}

The gas mass fractions computed as described above are tabulated in
Table~1 together with the input data.  In general there is close
correspondence between $f_g$ computed from $B$-band and $I$-band data.
This is encouraging since independent methods are used to estimate
$\Upsilon_*$ which is the more important factor in equation~2.
Variation in the two different determinations of $f_g$ give some
idea of the uncertainties therein.

\placetable{tab1}

In Fig.~7 we plot the gas fractions against the $B$-band
properties of the galaxies.  A strong relation between gas fraction and
total luminosity, disk surface brightness, and color is apparent.
The same relations with luminosity and surface brightness are also
apparent in the $I$-band data (Fig.~8).
There is no relation at all with the size of the optical disk.
Fits to the $B$-band data give
\begin{mathletters}
\begin{eqnarray}
f_g = 0.12(M_B+23) \\
f_g = 0.12(\csb-19) \\
f_g = -1.4[(B-V)-0.95]
\end{eqnarray}
\end{mathletters}
where the uncertainty in the slope is $\pm 0.07$ for $M_B$, $\pm 0.04$ for
\csb, and $\pm 0.5$ for color.  These relations are of course subject
to modest variation with
improvements in the relations governing $\Upsilon_*$ and $\eta$,
but it is clear that dimmer galaxies are systematically more gas rich.
The slopes of the $f_g$-$M_B$ and -\csb\ relations are indistinguishable.

\placefigure{fg}

\placefigure{fgI}

An important consequence of the systematic increase of gas fraction
with surface brightness is the end of the gas consumption paradox.
Since roughly equal numbers of galaxies exist at each surface brightness,
there are many galaxies which still retain substantial gas reservoirs
(see McGaugh\markcite{me} 1996a for the distribution functions).
Perhaps it is not surprising that the galaxies which are the most
prominent optically are those which have converted most of their
gas into stars.

\subsection{Correlations}

Correlation coefficients from a principle component analysis of the $B$-band
data are given in Table~2.  When reading this table, be sure to remember
the backwards convention in the definition of magnitudes.  This affects the
signs of the correlation coefficients:  ${\cal R} = -0.59$ between $M_B$ and $h$
means that brighter galaxies tend to be bigger, not smaller.

First note that $f_g$ is well correlated
with $M_{HI}/L$, from which it is derived.  This was not entirely a foregone
conclusion, since the relations for $\Upsilon_*$ and $\eta$ could
conceivably have offset this.  The use of these relations introduces
noise which reduces the correlation coefficient from unity.

\placetable{tab2}

Amongst independent variables,
the strongest correlation is between $M_{HI}/L$ and \csb.
This relation is more significant than either of the well known relations
between $M_{HI}/L$ and $M_B$ (Rao \& Briggs\markcite{RB} 1993;
Salpeter \& Hoffman\markcite{SH} 1996) or $M_{HI}/L$ and ${\cal T}$
(Roberts\markcite{RX} 1963), though these are also obvious in Table~2.
This strong $M_{HI}/L$-\csb\ relation translates into a strong one
between $f_g$ and \csb.
The central surface brightness of a galaxy is a good indicator of its
cumulative gas consumption and hence its evolutionary state.

The gas fraction is also correlated with $M_B$ and $B-V$.  Though both
are used in the calculation of $f_g$, this correlation is not an artifact.
It stems from the basic fact that there is relatively more 21 cm luminosity
in spirals of low optical luminosity, which also tend to be bluer.  The
relation with color goes in the sense that blue galaxies are more gas rich, as
one would expect.  Moreover, the same result is
obtained when color is not used in the calculation of $f_g$ with
the $I$-band data.

The data appear to occupy a fairly narrow structure in $f_g$-\csb-color-$M_B$
space.  In analogy to the Fundamental Plane for ellipticals,
this may indicate the existence of an ``Evolutionary Hypersurface''
constraining the evolution of spirals.  However, this is violated by
the extreme galaxy Malin~1 which is exceedingly gas rich for its
luminosity.  A more fundamental plane may be that in $f_g$-\csb-color space,
but this too may be modified by further selection effects (see below).

The gas fraction is not well correlated with size.
The evolutionary rate of spiral galaxies is apparently scale free.
Neither is $f_g$ correlated with Hubble type,
in spite of the $M_{HI}/L$-${\cal T}$ relation.  This is because
the dominant gas phase changes rapidly with Hubble type (Young \&
Knezek\markcite{H2H1} 1989), offsetting in the total gas fraction
the trend apparent in the atomic phase alone.

Among optical parameters, Hubble type depends most on $M_B$ and \csb.
Though it is not surprising that these are relevant to the morphological
appearance of a galaxy, ${\cal T}$
does not provide a clean measure of any physical
parameter (Naim\markcite{nnet} \etal 1995;
McGaugh\markcite{LSBmorph} 1995b; de Jong\markcite{dJt} 1995).
Type is not well correlated with size,
though there is some tendency for earlier types to be redder.

The linear size of a disk galaxy is not well correlated with anything
except absolute magnitude.  Absolute magnitude is appears to be
reasonably well correlated with surface brightness in the obvious sense
that higher surface brightness galaxies are brighter.  However, this
may be illusory, being the relation most subject to selection effects.
In particular, essentially all the low luminosity galaxies are low
surface brightness (recall Fig.~2).  This may or may not be representative
of reality.  If we restrict the sample to objects brighter than $-18$
where the $M_B$-\csb\ plane is more completely covered, this correlation
coefficient drops to ${\cal R} = 0.34$.  That between \csb\ and $h$ increases
to ${\cal R} = 0.44$ --- bigger galaxies tend to be lower surface brightness.
These are the only significant changes
in the correlation matrix when the sample is restricted to bright galaxies.

Optical galaxy selection depends jointly on $M_B$, \csb, and
$h$ (Disney \& Phillipps\markcite{DP} 1983;
McGaugh\markcite{MBS} \etal 1995a), so correlations between these
variables are very sensitive to sample selection and must be treated with
caution.  Correlations with other variables are not as directly subject
to selection effects, and can sensibly be analyzed as long as representative
data is available over a reasonable dynamic range.
In particular, correlations involving $f_g$ appear to be robust.

\subsection{Further Selection Effects}

We should nevertheless consider whether the relations derived above
are affected by selection effects which in the past helped conceal them.
We have addressed the problem of isophotal selection,
to the extent currently possible,
by specifically targeting low surface brightness galaxies.  This
increases the dynamic range available in \csb, but it
is obvious that such effects remain from the truncation of the data as
$\csb \rightarrow 25$.
The galaxies studied here are selected optically, but of course
a galaxy must have detected HI to perform this analysis.
The lack of low gas fraction, low surface brightness galaxies could therefore
be a further selection effect.

The relations appear continuous across samples, which would seem to argue
against strong selection effects in this sense.
The 21 cm detection rate is very high (80 \%) for low surface
brightness galaxies in the Schombert\markcite{lsbcat} \etal (1992) catalog,
so we do not seem to be missing a significant number of these predominantly
spiral galaxies.  However, the fact that these galaxies are very blue
(Fig.~3) may itself be a selection effect, since they were selected on blue
sensitive plates (see McGaugh\markcite{me} 1996a).  For this to be
important to Fig.~7, a large {\it additional\/} population of gas poor
red low surface brightness galaxies would also have to exist.

Other types of galaxies (\eg dE and dSph) are certainly
known to exist which have low surface brightnesses and low
gas fractions.  These particular types bear no apparent relation
to the larger low surface brightness spirals cataloged by
Schombert\markcite{lsbcat} \etal (1992), or to higher surface brightness
spirals.  Nevertheless, it may well be that the observed relations
represent only upper envelopes, and a substantial population of gas poor
low surface brightness galaxies remains to be discovered.
Indeed, such a population may be demanded by the rapid evolution of the
luminosity density (Dalcanton\markcite{D93} 1993; Lilly\markcite{CFRS}
\etal 1995 --- see \S 6).
The lack of gas rich, high surface brightness galaxies
is {\it not\/} a selection effect:  such objects would be
prominent both optically and in HI.

\section{Disk Evolution}

Irrespective of these caveats, clear evolutionary trends exist.
The lower surface brightness and lower luminosity spirals
that we know about have consumed
less of their gas and are less evolved than brighter systems.
This result is consistent with other lines of evidence from
star formation thresholds 
(van der Hulst \etal\markcite{vdH} 1993), abundances (McGaugh\markcite{OHme}
1994a; R\"onnback \& Bergvall\markcite{RB2} 1995), colors
(McGaugh \& Bothun\markcite{MB} 1994; R\"onnback \& Bergvall\markcite{RB1} 1994;
Bergvall \& R\"onnback\markcite{BR} 1995;
de Blok \etal\markcite{dB1} 1995) and environments (Bothun\markcite{B93} \etal
1993; Mo\markcite{mo} \etal 1994), all of which indicate that low surface
brightness galaxies evolve slowly, and perhaps form late.

On their own, Figs.~7 and 8 suggest a straightforward picture:
disk galaxies form and then evolve
at constant size, converting gas into stars and hence changing their luminosity
and surface brightness at the same rate.  This assures strong $f_g$-$M_B$ and
$f_g$-\csb\ relations with the same slope.  There need be no correlation
between $M_B$ and \csb\
since galaxies exist over a range of sizes.  Disk galaxy evolution thus seems
to proceed in an orderly fashion, without excessive amounts of merging, size
evolution, strong stochastic star bursts, or other exotic phenomena.

The evolutionary trajectory of a galaxy in Fig.~7 or 8 depends on the
evolution of stars it contains, and the rate at which those stars are made.
The star formation rate (SFR) is just the negative of the rate of
gas consumption: $\dot f_* = -\dot f_g$, at least until gas recycling becomes
important at low $f_g$.  The photometric evolution
of stars depends on many factors (\eg their metallicity) which lead to
a variety of complications, but the basic physics is well understood and
allows one to estimate the luminosity that will be produced by a given
mass of stars.  This gives the abscissa.  The ordinate follows from
the SFR which is traditionally parameterized as some convenient function of
time (\eg Guideroni \& Rocca-Volmerange\markcite{GRV} 1987).  For
declining SFRs in which most of the star formation occurred well in the
past, a commonly assumed form is
\begin{equation}
\dot f_* = -\dot f_g = \tau^{-1} e^{-t/\tau}.
\end{equation}
The SFR is normalized to the total baryonic mass of a given galaxy
so that $0 \le f_* \le 1$.
The $e$-folding time scale $\tau$ is typically several Gyr as
thought appropriate for early type galaxies.  For SFRs which are
constant (appropriate for late type spirals) or rising (as is perhaps
the case for irregular galaxies), a useful form is
\begin{equation}
\dot f_* = -\dot f_g = \frac{x+1}{\tau_g} \left(\frac{t}{\tau_g}\right)^{x}.
\end{equation}
In this case, $\tau_g$ is the time to total gas consumption.  This
is not quite the same as an $e$-folding time scale, but in both cases
$\tau^{-1}$ sets the evolutionary rate.

We do not delude ourselves that either of these simple SFRs accurately
represent what really occurs in any given galaxy.  They do nevertheless
provide a very useful way of parameterizing the range of possible average
star formation histories.  This simple analytic approach does enable one
to infer several interesting constraints on the more complicated reality
when a large dynamic range in $f_g$ is available.

Integration of these SFRs allows us to define several quantities of
interest (and also illustrates why these forms are simple and popular).
The age of a galaxy, $T_G$, we will take to mean the time since
star formation commenced.  This is the obvious working definition of
galaxy formation in this context.  It does not, of course, specify
precisely how a galaxy formed.  Star formation might well begin while
the collapse of gas is ongoing, or might have begun in separate sub units
which subsequently merge into the single entity observed today.
It is convenient to imagine the case in which the time scale for such
formation events is short compared to the evolutionary time scale $\tau$
so that formation is contemporaneous with the onset of star formation
and $T_G$ is the true age of the galaxy for all practical purposes.
This need not be the case, so one should keep in mind that $T_G$
really measures the time since the onset of star formation.

Another interesting quantity is the mean age of the stars,
$\langle T_* \rangle$.  It is often possible to vary the shape of the
SFR so as to arrive at any desired $f_g$ with a degenerate set of
ages $T_G$ and evolutionary rates $\tau$.  However, otherwise plausible
combinations of these two sometimes have implausible consequences for
$\langle T_* \rangle$.  These and other quantities and limits of interest
are given in Table~(3).

\placetable{tab3}

Equipped with this formalism, we are nearly ready to interpret the positions
of galaxies in Fig.~8.  The value of $f_g$ comes from stipulating an SFR,
but the precise trajectory a galaxy follows in luminosity depends on the
details of the stars for which a population model is required.  This gives
the variation of $\Upsilon_*$ with time.  The amount of light per unit mass
produced by a single generation of stars declines rapidly as luminous, massive
stars evolve and die.  However, if star formation is ongoing for any period
of time, this effect is offset by the birth of new stars.  To estimate
the interpay of these effects, we need to choose a specific population model.
For now, we employ the models of Guideroni \& Rocca-Volmerange\markcite{GRV}
(1987), briefly described in Table~(4).  This particular choice is made
for the simple reason that they published the full range of $f_g$
for an interesting variety of SFRs.  Various improvements in modeling have
since been made, but no uniform set of models is readily available over the
entire dynamic range we need.  Moreover, we are trying here to avoid as much as
possible the serious astronomical problem of model dependency; specific
models are adopted only to illustrate the various issues.  Compared to
the dynamic range in $f_g$, stellar populations experience fairly mild
evolution of $\Upsilon_*$ in all plausible models.  So while the detailed
shape of evolutionary tracks is highly model dependent, the basic gist is
not.

For example, Fig.~8 by itself does not represent a direct evolutionary sequence.
Galaxies do not start at faint $M$ and \csb\ and simply
brighten along the observed sequence.  The slope of the observed relations
is an order of magnitude different from what is expected from evolutionary
models for the rate of change of absolute magnitude ($\Delta m$) with gas
fraction ($\Delta f_g$).  Instead of the slope observed in equation~(6),
$\frac{\Delta m}{\Delta f_g} = 8$, a very broad range of models
evolve slowly in luminosity compared to the rate of
gas consumption with $-2 < \frac{\Delta m}{\Delta f_g} < 2$ (Fig.~9).
This difference between observed and model slopes requires a systematic
variation in the evolutionary rate with surface brightness.
Progressively dimmer galaxies either evolve more slowly or are younger,
or some combination thereof.

The isochrones of the models of Guideroni \& Rocca-Volmerange\markcite{GRV}
(1987) are roughly parallel to the observed $\frac{\Delta m}{\Delta f_g}$ slope
for ages $T_G > 10$ Gyr.  This is achieved by systematically changing the
gas consumption time scale with Hubble type (Table 4).  Since
very little of these data were available when the models were constructed,
this is a success of the standard picture of galaxy evolution
the models represent.  On the other hand, the correlation between
$M_{\rm HI}/L$ and type was known, and the models were constructed to
match that data.

\placefigure{Tracks}

From Fig.~9 one can read off the combined age-SFR of any given galaxy.
If one can be constrained by other information like colors, the other
follows.  Type is correlated with surface brightness (Table~2;
de Blok\markcite{dB1} \etal 1995; de Jong\markcite{dJ3} 1996), so it is
possible to make all spirals the same old ($\sim 15$ Gyr) age and
satisfy the observed gas fractions simply by varying the time scale
$\tau$.  The variation must be quite systematic with surface brightness,
and so large that one has to change the functional form of the SFR to
accommodate it.  (In principle, one could retain an exponential SFR by
allowing $\tau \rightarrow \infty$ and then becoming negative.)

The spread of the data is large compared to the separation
of the isochrones.  This may indicate a large dispersion in age.
More likely, it simply results from an
overlap of the range of baryonic masses of any given SFR type.
That is, a particular SFR type like Sd may be experienced
by galaxies of a range of masses, not just one particular one
as drawn in Fig.~9.  Because there is inevitably overlap of
this sort we do not attribute any special significance to the kink
in the model grid for Sa and Sb SFRs at low $f_g$ --- the overlap
would wash it out, even if the models were credible at this level
of detail.  The lack of any corresponding feature in the
data might indicate that $\tau$ does not vary as widely as assumed
in the models, so that for spiral disks deviations from a constant
SFR are not enormous ($\tau \gtrsim 8$ Gyr).

\placetable{tab4}

The standard picture of galaxy evolution represented by the models of
Guideroni \& Rocca-Volmerange\markcite{GRV} (1987) must attribute the
observed change of $f_g$ with surface brightness to variations in $\tau$
because $T_G$ is taken to be the same for all galaxies.  Once this artificial
boundary condition is imposed, there is no choice.  It is not obvious that
this is really a sensible assumption.  If galaxy formation occurs from the
gravitational collapse of primordial fluctuations in the density field,
then the epoch of collapse of any given object depends on the density contrast
of the fluctuation from which it arises.  Most of the observational
evidence (McGaugh 1992\markcite{mythesis}; 1996b\markcite{IAU}) suggests
that this occurs in the obvious way:  large density contrasts result in
high surface brightness galaxies, and small density contrasts result in low
surface brightness galaxies.  If this is indeed the case, one expects
lower surface brightness galaxies to be younger as inferred from their colors
(McGaugh \& Bothun\markcite{MB} 1994; de Blok\markcite{dB1} \etal 1995).
The absolute magnitude of the age difference may or may not be large;
it is very difficult to put an absolute number on.  From the colors,
the relative shift typically appears to be on the order of a few
(3 - 5) Gyr.  If high surface brightness spirals are 13 or 14 Gyr old,
low surface brightness disks are roughly 8 or 10 Gyr old.

To examine this possibility in the context of the gas fraction data,
consider as a trivial limiting case a scenario in which all spiral disks have
the same shape star formation history.  The evolutionary trajectories
would all look the same in Fig.~9, and (by choice) isochrones
would be horizontal lines.  Instead of a systematic variation in the
evolutionary time scale, real age differences would be required to explain
the trend of the data.  For the case of constant star formation at the
same relative rate (the Sd track), bright galaxies with $f_g \approx 0.3$ would
have ages $T_G \approx 17$~Gyr, while the dim ones with $f_g \approx 0.6$ would
be $T_G \approx 10$~Gyr old.

Either of these scenarios seem plausible, and reality is presumably
more complicated --- both age and evolutionary rate may play a role.
Gas fractions alone do not allow us to distinguish which factor dominates.
They do show that clear differences exist. This
may help lift the degeneracy that has plagued analyses of
colors alone.  If so, it may become possible to
age-date spiral galaxies.

The distinction between a galaxy which is gas rich today because it
formed early and evolved slowly or because it formed late is not clear.
Some limit can be placed on these extreme cases by the plausibility of
the required SFR and mean stellar age.
If we insist that all galaxies have the same age,
the only way to have very gas rich blue galaxies
($f_g > 0.4$) at $z=0$ is to have star formation rates which increase
strongly with time.  These retain much gas today simply because very
little happened for the first half of a Hubble time.  For the Im track in
Fig.~9, only 10\% of the gas is consumed in the first 9 Gyr.
The distinction between this and a galaxy which
forms late is inobvious, and returns us to the thorny issue of the
definition of galaxy formation.  Perhaps a lengthy collapse by gradual
dissipation and a slow evolution are intimately related.

For a more quantitative illustration, consider the conservative
case of a declining star formation rate.  Retaining a gas fraction
greater than what is consumed in one $e$-folding ($f_g > e^{-1} = 0.37$,
corresponding to $\csb > 22$) requires
an evolutionary time scale $\tau > T_U$ greater than the age of the
universe, with a mean stellar age $\langle T_* \rangle < 0.6 T_U$.
This follows simply from requiring that the galaxy be younger than
the universe: $T_G < T_U$.  If we adopt a rising
star formation history as suggested by the Im model in Fig.~9,
the evolutionary time scale can be very (but not arbitrarily) long.
This is only achieved at the expense of making the stars very young:
$\langle T_* \rangle < \onequarter T_U$, \ie $\langle T_* \rangle =
3$ Gyr for $T_G = 12$ Gyr.  Either way,
most of the star formation in very gas rich galaxies must be weighted
towards more recent epochs (\cf McGaugh \& Bothun\markcite{MB} 1994),
irrespective of whether the galaxy formed early and did nothing, or formed
late.  Indeed, galaxies with $f_g > 0.5$ have yet to realize most of
their potential for star formation, which must be weighted towards
{\it future\/} epochs.

\section{Conundrum}

The picture which emerges thus far seems fairly sensible, confirming and
extending established wisdom about the evolution of local galaxies.
Related information can be inferred from deep redshift surveys, and
considerable progress has been made in measuring the properties of
galaxies over an interesting range of redshift.  These studies indicate
a large amount of evolution in both field (Lilly\markcite{CFRS}
\etal 1995; Ellis\markcite{ECBHG} \etal 1996) and cluster
(Schade\markcite{CNOC} \etal 1996) galaxies.  High resolution imaging with
{\sl HST\/} shows that most of this evolution is due to morphologically
late type galaxies (Glazebrook\markcite{kgb} \etal 1995;
Driver\markcite{driver} \etal 1995, Abraham\markcite{HDF} \etal 1996).
This is consistent with the evolution of the luminosity function
constructed from deep surveys:  red galaxies appear to evolve little,
while blue ones have been fading very rapidly (Lilly\markcite{CFRS}
\etal 1995; Ellis\markcite{ECBHG} \etal 1996).

There is good evidence for direct evolution of the surface brightnesses
of bright disks (Schade \etal 1995\markcite{schade}, 1996\markcite{CNOC}).
These appear to have surface brightnesses which are $\sim 1\sb$ brighter at
$z \approx 0.5$ than at $z = 0$.  This is perfectly consistent with Fig.~9:
these galaxies were presumably more gas rich at $z = 0.5$ and have since
evolved along a track like that plotted for early-type (Sa-Sc) spirals.

The local and high redshift approaches to the evolution of early type spirals
therefore give nicely consistent results.  The same can not be said for late
types.  Many studies now indicate that the excess faint blue galaxies are
predominantly late types which are evolving rapidly.  If one looks locally
for faint, blue, late type galaxies, one finds low surface brightness galaxies.
Indeed, the similarity between the physical properties of these two
populations led to the suggestion that one might account for the other
(McGaugh\markcite{FBGs} 1994b).  This similarity now extends to morphology:
these galaxies literally look alike (\cf Dalcanton \& Shectman\markcite{chain}
1996).  However, a simple one to one connection is not sufficient: although
deep surveys are more sensitive to low surface brightness galaxies than
local ones, they are still selected by flux.  Magnitude limited surveys
are always numerically dominated by the highest surface brightness galaxies
regardless of how common low surface brightness galaxies actually are
(McGaugh\markcite{MBS} \etal 1995a), so for most plausible scenarios
low surface brightness galaxies do not make a significant contribution
to the faint counts (Ferguson \& McGaugh\markcite{harry} 1995;
McLeod \& Rieke\markcite{MLR} 1995) unless they evolve substantially.

It is tempting to conclude
that this is exactly what happens:  late type blue galaxies evolve rapidly,
fading to become low surface brightness by the present epoch (Phillipps \&
Driver\markcite{PD} 1995).  While sensible in the context of the high
redshift data, it makes no sense for the known local galaxies.
Fading scenarios for the non-dwarf low surface brightness galaxies
in the present sample are strongly excluded by both the colors (McGaugh \&
Bothun\markcite{MB} 1994; Fig.~3) and by the gas fractions.
In order to have a strongly fading trajectory (Sa or even more extreme)
in Fig.~9 and retain the
high observed gas fractions, such galaxies would have to be extremely
young:  $T_G < 2$ Gyr for $\tau = 3$ Gyr and $f_g > 0.5$.
They would not have even formed yet at the redshifts where
the inferred evolution occurs.

Indeed, the evolutionary histories discussed
in \S 5 are precisely the opposite:  late type, low
surface brightness galaxies evolve slowly, not rapidly.
Photometrically, they evolve at roughly constant luminosity
or even {\it brighten\/}
with time.  We are left with a puzzling conundrum: there exist two populations,
the faint blue galaxies at high redshift and low surface brightness galaxies
locally, which have very similar colors, clustering properties, and
morphologies:  they literally look alike, yet they can not be the same.

Perhaps the picture which emerges from the deep survey work is
deceptively simple.  Dividing the luminosity function into bins by
color (Lilly\markcite{CFRS} \etal 1995) or spectral type
(Heyl\markcite{jeremy} \etal 1996) gives the appearance of a slowly
or even non-evolving red population and a rapidly evolving blue one.
Yet one expects rather more evolutionary fading for red galaxies,
consistent with the amount of surface brightness evolution observed
by Schade \etal (1996).  Perhaps there is some cross-over between
color bins.  This is
expected at some level, as even early type galaxies should have been
blue at some point early in their evolution.

It is not obvious that cross-over is adequate to explain the evolution
of early types, but it may well be (Pozzetti\markcite{Pozz} \etal 1996).
A more serious problem is understanding the rapid evolution of the
blue luminosity density attributed to late types.  Since the faint blue
galaxies we do know about locally seem unable to explain this, perhaps
the inference of a rapidly evolving dwarf population is correct and
there exists locally an additional remnant population
(Phillipps \& Driver\markcite{PD} 1995; Babul \& Ferguson\markcite{BF}
1996).  Such a population would presumably now have faded and become
red, low surface brightness, and gas poor.  This is precisely the part
of the $f_g$-\csb\ plane not constrained by observation.  A very large
population of such objects would be required, but it is quite possible
that such a population would have remained undetected so
far (McGaugh\markcite{me} 1996a), especially if the typical object is
intrinsically small as indicated by deep {\sl HST\/} imaging (Im\markcite{IM}
\etal 1995).  This picture might be considered attractive for avoiding the
reverse of the gas consumption paradox:  if gas rich galaxies exist as
well as those near to depleting their gas supply, why are there no galaxies
well past this point?  The unavoidable consequence of such a scenario is that
our knowledge of the local galaxy population remains glaringly incomplete.

\section{Conclusions}

Examination of the gas mass fractions of spiral galaxies over a large range
in luminosity and surface brightness provides a number of insights
into galaxy evolution.
\begin{enumerate}
\item The fraction of baryonic mass in spiral galaxies which has been converted
from gas into stars is correlated with luminosity, surface brightness, and
color.  It is not correlated with the scale length of the disk.
\item The strongest correlation is between gas content and the central surface
brightness of the disk.  The surface brightness of a disk is a fundamental
parameter and a good indicator of the evolutionary status of a galaxy.
 Dim galaxies evolve slowly, have relatively young stellar populations,
and may have formed late.
\item Since luminosity traces baryonic mass and surface brightness traces
surface mass density, the strength of the $f_g$-\csb\ relation suggests that
surface density is at least as important as total mass in determining
the evolution of a galaxy.
\item The $f_g$-$M_B$ and $f_g$-\csb\ relations have the same slope.
Any size evolution would cause surface brightness evolution over and above
that due to luminosity evolution, leading to a difference in these slopes.
The lack of such a difference indicates that spirals evolve at roughly fixed
size.
\item The large number of gas rich low surface brightness galaxies resolves
the gas consumption paradox.
\item The populations of local low surface brightness galaxies and high
redshift faint blue galaxies are physically and morphologically similar
yet grossly different in the way they evolve.  This poses a difficult
conundrum.  One possible solution is an additional population of local,
red, gas poor low surface brightness galaxies.
\end{enumerate}

There is no reason to believe that we have reached the faintest limits
of surface brightness, or that yet more gas rich galaxies do not exist.
Given the barrier imposed by isophotal selection limits,
there could certainly be more very low surface brightness
galaxies that have yet to be discovered.  Some of
these could be even more gas rich (by extrapolation of the observed trends)
and others might be gas poor (\S 6).

Several further lines of work are suggested.
\begin{enumerate}
\item Further surveys for low surface brightness galaxies, both to
deeper isophotal limits and in redder passbands.  Blind 21 cm surveys with
optical follow-up to measure surface brightnesses would also be interesting.
\item Detailed modeling of individual galaxies to match colors, metallicities,
star formation rates and gas mass fractions.
\item Examining these properties on a local as well as global basis.
\end{enumerate}
These projects would address some of the puzzles posed, and further advance
our understanding of galaxy evolution.

In essence, we must understand how low surface brightness
galaxies can remain gas rich until the present epoch.
Either quiescence or youth, or some combination thereof, can be invoked.
If quiescence is primarily responsible, it must be possible to suppress
large scale star formation in gas disks for many Gyr, perhaps through critical
density dependent phenomena (\eg Kennicutt\markcite{K89} 1989).
If youth is a factor, the existence of such unevolved galaxies suggests
that the process of galaxy formation is ongoing, or has ended only recently.

\acknowledgments  We are grateful to J. M. van der Hulst and L.B. van den Hoek
for helpful discussions, and thank the referee for suggesting
a more expanded discussion.


\begin{deluxetable}{lccrcrccccc}
\tablewidth{0pt}
\tablecaption{Data \label{tab1}}
\tablehead{\colhead{Galaxy} & \colhead{$M_B$} & \colhead{\csb} & \colhead{$h$}
& \colhead{$B-V$} & \colhead{${\cal T}$} & \colhead{$\frac{M_{\rm HI}}{L_B}$} &
\colhead{$f_g^B$} & \colhead{$\frac{M_{\rm HI}}{L_I}$} & \colhead{$f_g^I$} &
\colhead{$B-I$} }
\startdata
DDO 142 & $-$20.61 & 22.23 &   3.7 & 0.68   & 9 & 0.66  & 0.33  & \nodata&\nodata&\nodata \nl
F415--3 & $-$16.48 & 23.87 &   1.6 & 0.52   & 9 & 1.71  & 0.64  & 1.91   & 0.68  & 1.23 \nl
F469--2 & $-$17.39 & 24.85 &   3.8 & 0.43   & 9 & 0.95  & 0.56  & 0.93   & 0.51  & 1.37 \nl
F530--3 & $-$18.77 & 23.85 &   4.9 & 0.64   & 5 & 0.53  & 0.44  & 0.46   & 0.49  & 1.49 \nl
F561--1 & $-$17.83 & 23.28 &   3.6 & 0.59   & 9 & 0.68  & 0.38  & 0.76   & 0.46  & 1.22 \nl
F563--1 & $-$17.33 & 24.00 &   4.2 & 0.65   & 9 & 2.05  & 0.62  & \nodata&\nodata&\nodata \nl
F563--V1 & $-$16.36 & 24.29 &  2.4 & 0.56   &10 & 0.90  & 0.47  & 1.06   & 0.55  & 1.18 \nl
F563--V2 & $-$18.21 & 22.22 &  2.1 & 0.36   &10 & 0.76  & 0.55  & 0.62   & 0.42  & 1.57 \nl
F565--V2 & $-$15.41 & 24.73 &  2.6 & 0.53   &10 & 2.62  & 0.73  & \nodata&\nodata&\nodata \nl
F567--2 & $-$17.38 & 24.41  &  5.7 & 0.67   & 9 & 1.56  & 0.54  & 2.32   & 0.72  & 0.92 \nl
F568--1 & $-$18.12 & 23.76  &  5.3 & 0.62   & 5 & 1.45  & 0.69  & 1.49   & 0.75  & 1.32 \nl
F568--3 & $-$18.31 & 23.07  &  4.0 & 0.55   & 7 & 0.85  & 0.51  & 0.90   & 0.55  & 1.29 \nl
F568--6 & $-$21.57 & 23.60  & 21.1 & 0.63   & 5 & 0.44  & 0.32  & 0.19   & 0.28  & 2.26 \nl
F568--V1 & $-$17.88 & 23.29 &  3.2 & 0.51   & 8 & 1.11  & 0.56  & 1.26   & 0.60  & 1.21 \nl
F571--5 & $-$17.14 & 23.65  &  2.9 & 0.34   & 9 & 1.55  & 0.73  & \nodata&\nodata&\nodata \nl
F571--V1 & $-$17.04 & 23.99 &  3.2 & 0.53   & 8 & 1.13  & 0.56  & \nodata&\nodata&\nodata \nl
F574--2 & $-$17.64 & 24.33  &  6.0 & 0.63   & 9 & 0.92  & 0.44  & \nodata&\nodata&\nodata \nl
F577--V1 & $-$18.21 & 23.94 &  5.7 & 0.38   & 7 & 0.83  & 0.61  & \nodata&\nodata&\nodata \nl
F611--1 & $-$15.76 & 24.65  &  2.0 & 0.44   &10 & 1.07  & 0.58  & 0.99   & 0.53  & 1.44 \nl
F615--1 & $-$17.54 & 23.36  &  2.8 & 0.51   &10 & 0.61  & 0.40  & 0.58   & 0.40  & 1.40 \nl
F746--1 & $-$19.42 & 23.24  &  4.0 & 0.65   &10 & 0.73  & 0.37  & 0.71   & 0.45  & 1.38 \nl
Malin 1 & $-$22.30 & 26.60  & 73.0 & 0.90   &\nodata & 3\phm{.}\phn\phn & 0.75 & \nodata&\nodata&\nodata \nl
NGC 3913  & $-$19.05 & 22.60 & 2.1 & 0.60   & 7 & 0.29  & 0.24  & \nodata&\nodata&\nodata \nl
NGC 4411A & $-$19.13 & 22.07 & 1.7 & 0.68   & 8 & 0.34  & 0.21  & \nodata&\nodata&\nodata \nl
NGC 4411B & $-$19.65 & 22.07 & 2.2 & 0.56   & 8 & 0.33  & 0.25  & \nodata&\nodata&\nodata \nl
NGC 5774 & $-$20.55 & 22.35  & 3.8 & 0.54   & 7 & 0.86  & 0.51  & \nodata&\nodata&\nodata \nl
UGC 89  & $-$21.50 & 22.07   & 8.7 & 0.74   & 1 & 0.13  & 0.26  & 0.09   & 0.29  & 1.67 \nl
UGC 93  & $-$20.06 & 22.33   & 7.0 & 0.64   & 8 & 0.78  & 0.41  & 0.54   & 0.38  & 1.75 \nl
UGC 128 & $-$18.78 & 24.22   & 6.8 & 0.51   & 6 & 1.24  & 0.66  & 1.57   & 0.72  & 1.09 \nl
UGC 334 & $-$18.99 & 23.36   & 6.6 & 0.90   & 9 & 0.92  & 0.32  & 1.02   & 0.51  & 1.24 \nl
UGC 438 & $-$21.10 & 20.45   & 4.2 & 0.73   & 5 & 0.15  & 0.16  & 0.09   & 0.15  & 1.89 \nl
UGC 463 & $-$20.74 & 20.76   & 4.0 & \nodata& 5 & 0.17  &\nodata&0.11   & 0.18  & 1.72 \nl
UGC 490 & $-$20.57 & 21.47   & 5.6 & 0.86   & 5 & 0.63  & 0.38  & 0.34   & 0.41  & 2.00 \nl
UGC 508 & $-$21.56 & 22.05   & 9.5 & 0.92   & 2 & 0.17  & 0.22  & 0.09   & 0.24  & 2.01 \nl
UGC 1230 & $-$18.33 & 23.36  & 5.3 & 0.52   & 5 & 1.72  & 0.77  & 1.98   & 0.80  & 1.20 \nl
UGC 1305 & $-$20.30 & 22.02  & 6.0 & 0.93   & 4 & 0.16  & 0.15  & 0.07   & 0.14  & 2.24 \nl
UGC 1551 & $-$19.46 & 22.47  & 4.6 & 0.62   & 8 & 0.49  & 0.31  & 0.43   & 0.31  & 1.48 \nl
UGC 1577 & $-$20.64 & 22.44  & 7.7 & 0.93   & 4 & 0.38  & 0.30  & 0.22   & 0.34  & 1.95 \nl
UGC 1719 & $-$21.35 & 22.45  &11.6 & 0.82   & 3 & 0.27  & 0.31  & 0.16   & 0.31  & 1.90 \nl
UGC 1792 & $-$20.36 & 21.65  & 5.6 & 0.85   & 5 & 0.61  & 0.38  & 0.46   & 0.44  & 1.65 \nl
UGC 2064 & $-$19.97 & 22.28  & 5.6 & 0.88   & 4 & 0.47  & 0.36  & 0.25   & 0.35  & 2.01 \nl
UGC 2124 & $-$19.32 & 22.34  & 3.9 & 0.98   & 1 & 0.63  & 0.56  & 0.22   & 0.51  & 2.50 \nl
UGC 2125 & $-$20.37 & 23.20  & 9.1 & 0.85   & 5 & 0.54  & 0.35  & 0.37   & 0.39  & 1.74 \nl
UGC 2197 & $-$19.95 & 22.57  & 6.1 & 0.73   & 6 & 0.64  & 0.39  & 0.48   & 0.38  & 1.66 \nl
UGC 2368 & $-$19.89 & 23.67  & 6.6 & 0.70   & 3 & 0.70  & 0.58  & 0.48   & 0.55  & 1.76 \nl
UGC 3066 & $-$19.96 & 22.03  & 4.3 & 0.73   & 7 & 0.73  & 0.38  & 0.44   & 0.31  & 1.88 \nl
UGC 3080 & $-$19.55 & 21.99  & 3.8 & 0.65   & 5 & 0.67  & 0.49  & 0.47   & 0.46  & 1.72 \nl
UGC 4126 & $-$20.89 & 21.83  & 6.9 & 0.63   & 3 & 0.21  & 0.32  & 0.24   & 0.30  & 1.62 \nl
UGC 4256 & $-$21.09 & 21.18  & 5.9 & 0.49   & 5 & 0.32  & 0.40  & 0.13   & 0.28  & 1.82 \nl
UGC 4308 & $-$20.29 & 21.34  & 4.5 & 0.59   & 5 & 0.35  & 0.37  & 0.26   & 0.33  & 1.68 \nl
UGC 4368 & $-$20.01 & 21.52  & 4.3 & 0.57   & 6 & 0.55  & 0.44  & 0.38   & 0.38  & 1.76 \nl
UGC 4422 & $-$21.02 & 22.04  & 8.5 & 0.47   & 5 & 0.26  & 0.38  & 0.17   & 0.27  & 1.77 \nl
UGC 4458 & $-$21.02 & 21.72  & 5.9 & 0.84   & 1 & 0.21  & 0.32  & 0.12   & 0.34  & 1.96 \nl
UGC 4841 & $-$19.84 & 22.80  & 3.3 & 0.67   & 7 & 0.54  & 0.34  & \nodata&\nodata&\nodata \nl
UGC 4922 & $-$19.88 & 23.13  & 4.0 & 0.57   & 7 & 0.60  & 0.41  & \nodata&\nodata&\nodata \nl
UGC 5303 & $-$19.05 & 21.32  & 3.1 & 0.65   & 5 & 0.24  & 0.26  & 0.17   & 0.26  & 1.69 \nl
UGC 5510 & $-$18.77 & 20.66  & 1.6 & \nodata& 4 & 0.38  &\nodata&0.37   & 0.47  & 1.38 \nl
UGC 5554 & $-$18.15 & 20.98  & 1.4 & \nodata& 1 & 0.12  &\nodata&0.08   & 0.26  & 1.82 \nl
UGC 5633 & $-$17.04 & 23.13  & 2.3 & 0.61   & 8 & 1.03  & 0.49  & \nodata&\nodata&\nodata \nl
UGC 5675 & $-$16.31 & 23.77  & 2.0 & \nodata& 9 & \nodata&\nodata&0.79   & 0.47  & 1.80 \nl
UGC 5709 & $-$19.77 & 23.60  & 5.4 & 0.48   & 5 & 0.52  & 0.52  & 0.33   & 0.41  & 1.83 \nl
UGC 5842 & $-$18.54 & 21.44  & 2.3 & \nodata& 6 & \nodata&\nodata&0.21   & 0.25  & 1.49 \nl
UGC 6151 & $-$17.49 & 23.26  & 2.8 & 0.51   & 9 & 0.64  & 0.41  & 0.61   & 0.41  & 1.41 \nl
UGC 6277 & $-$19.08 & 20.88  & 2.0 & \nodata& 5 & 0.25  &\nodata&0.18   & 0.27  & 1.69 \nl
UGC 6445 & $-$18.39 & 20.55  & 1.2 & 0.64   & 4 & 0.59  & 0.51  & 0.35   & 0.47  & 1.91 \nl
UGC 6453 & $-$18.38 & 20.91  & 1.5 & 0.89   & 4 & 0.58  & 0.40  & 0.47   & 0.54  & 1.56 \nl
UGC 6614 & $-$20.29 & 24.85  &15.9 & 0.72   & 5 & 0.97  & 0.55  & 0.39   & 0.45  & 2.32 \nl
UGC 6693 & $-$21.33 & 21.60  & 8.3 & 0.69   & 4 & 0.43  & 0.42  & 0.29   & 0.43  & 1.76 \nl
UGC 6746 & $-$21.46 & 21.38  & 7.8 & 0.81   & 1 & 0.14  & 0.29  & \nodata&\nodata&\nodata \nl
UGC 6922 & $-$17.76 & 23.52  & 1.8 & 0.61   &10 & 0.51  & 0.31  & \nodata&\nodata&\nodata \nl
UGC 6956 & $-$17.69 & 23.70  & 1.9 & 0.56   & 9 & 0.52  & 0.33  & \nodata&\nodata&\nodata \nl
UGC 6983 & $-$19.43 & 22.77  & 2.7 & 0.54   & 6 & 0.78  & 0.54  & \nodata&\nodata&\nodata \nl
UGC 7169 & $-$19.67 & 20.11  & 1.8 & 0.40   & 5 & 0.36  & 0.48  & \nodata&\nodata&\nodata \nl
UGC 7315 & $-$17.67 & 19.99  & 0.8 & 0.69   & 4 & 0.06  & 0.09  & 0.03   & 0.08  & 2.02 \nl
UGC 7450 & $-$21.43 & 21.17  & 6.4 & 0.59   & 4 & 0.10  & 0.17  & 0.08   & 0.17  & 1.58 \nl
UGC 7523 & $-$18.58 & 21.49  & 2.0 & 0.70   & 3 & 0.05  & 0.09  & 0.03   & 0.09  & 1.93 \nl
UGC 7557 & $-$19.12 & 22.88  & 2.5 & 0.41   &10 & 0.78  & 0.52  & \nodata&\nodata&\nodata \nl
UGC 7685 & $-$19.42 & 22.89  & 2.9 & 0.65   &10 & 0.72  & 0.37  & \nodata&\nodata&\nodata \nl
UGC 7901 & $-$18.62 & 20.07  & 1.2 & 0.63   & 5 & 0.34  & 0.34  & 0.23   & 0.31  & 1.77 \nl
UGC 7911 & $-$19.36 & 22.92  & 2.8 & 0.48   & 9 & 0.42  & 0.33  & \nodata&\nodata&\nodata \nl
UGC 8279 & $-$19.66 & 20.52  & 2.3 & \nodata& 5 & 0.27  &\nodata&0.19   & 0.29  & 1.68 \nl
UGC 8289 & $-$20.91 & 21.80  & 7.1 & \nodata& 4 & 0.32  &\nodata&0.27   & 0.41  & 1.50 \nl
UGC 8865 & $-$19.93 & 21.89  & 5.0 & \nodata& 2 & 0.22  &\nodata&0.15   & 0.37  & 1.72 \nl
UGC 9024 & $-$18.73 & 24.66  & 7.5 & \nodata& 9 & 0.82  &\nodata&0.89   & 0.50  & 1.25 \nl
UGC 9061 & $-$22.27 & 22.63  &21.9 & \nodata& 4 & 0.33  &\nodata&0.19   & 0.32  & 1.95 \nl
UGC 9481 & $-$20.22 & 21.22  & 4.3 & \nodata& 4 & 0.52  &\nodata&0.42   & 0.52  & 1.57 \nl
UGC 9926 & $-$20.09 & 20.13  & 2.3 & 0.63   & 5 & 0.19  & 0.22  & \nodata&\nodata&\nodata \nl
UGC 9943 & $-$19.92 & 20.40  & 2.6 & 0.67   & 5 & 0.21  & 0.23  & 0.14   & 0.22  & 1.80 \nl
UGC 10083 & $-$19.79 & 21.51 & 2.9 & 0.42   & 2 & 0.27  & 0.55  & 0.27   & 0.49  & 1.34 \nl
UGC 10437 & $-$18.45 & 24.03 & 5.9 & 0.29   & 7 & 1.10  & 0.73  & 1.05   & 0.59  & 1.40 \nl
UGC 10584 & $-$21.20 & 21.77 & 8.3 & 0.61   & 5 & 0.55  & 0.47  & 0.38   & 0.44  & 1.76 \nl
UGC 11628 & $-$21.36 & 22.27 &11.0 & 0.96   & 2 & 0.26  & 0.29  & 0.18   & 0.36  & 1.76 \nl
UGC 11708 & $-$20.65 & 21.51 & 5.0 & 0.66   & 5 & 0.27  & 0.25  & 0.16   & 0.21  & 1.87 \nl
UGC 11868 & $-$18.56 & 22.65 & 1.7 & 0.74   &10 & 0.42  & 0.22  & \nodata&\nodata&\nodata \nl
UGC 11872 & $-$19.46 & 20.46 & 1.5 & 0.84   & 3 & 0.21  & 0.24  & 0.11   & 0.23  & 2.03 \nl
UGC 12151 & $-$17.76 & 23.27 & 3.1 & 0.55   &10 & 1.45  & 0.59  & 1.47   & 0.61  & 1.33 \nl
UGC 12343 & $-$20.93 & 21.95 & 8.3 & 0.98   & 5 & 0.26  & 0.18  & 0.14   & 0.22  & 1.98 \nl
UGC 12379 & $-$21.37 & 21.99 & 8.0 & 0.80   & 4 & 0.42  & 0.36  & 0.20   & 0.31  & 2.14 \nl
UGC 12391 & $-$20.47 & 21.50 & 5.1 & 0.64   & 5 & 0.73  & 0.52  & 0.48   & 0.48  & 1.79 \nl
UGC 12511 & $-$19.76 & 22.46 & 5.8 & 0.72   & 6 & 1.52  & 0.61  & 0.76   & 0.55  & 2.09 \nl
UGC 12614 & $-$20.91 & 20.94 & 5.1 & 0.73   & 5 & 0.20  & 0.20  & 0.14   & 0.22  & 1.71 \nl
UGC 12638 & $-$20.72 & 22.17 & 9.1 & 0.75   & 5 & 0.30  & 0.25  & 0.21   & 0.26  & 1.76 \nl
UGC 12654 & $-$20.17 & 21.76 & 5.3 & 0.78   & 4 & 0.44  & 0.38  & 0.31   & 0.43  & 1.72 \nl
UGC 12695 & $-$18.92 & 24.59 & 8.4 & 0.37   & 9 & 1.28  & 0.67  & 1.31   & 0.60  & 1.32 \nl
UGC 12754 & $-$18.74 & 21.81 & 3.1 & 0.54   & 6 & 0.37  & 0.35  & 0.36   & 0.36  & 1.37 \nl
UGC 12776 & $-$21.25 & 23.39 &20.3 & \nodata& 3 & 0.38  &\nodata&0.29   & 0.45  & 1.62 \nl
UGC 12808 & $-$21.41 & 20.31 & 3.9 & 0.70   & 3 & 0.09  & 0.16  & 0.05   & 0.15  & 1.85 \nl
UGC 12845 & $-$20.19 & 22.69 & 8.0 & 0.89   & 7 & 0.79  & 0.33  & 0.56   & 0.42  & 1.72 \nl
\enddata
\end{deluxetable}

\begin{deluxetable}{lcccccc}
\tablewidth{0pt}
\tablecaption{Correlation Matrix \label{tab2}}
\tablehead{
& \colhead{\phs $f_g$} & \colhead{\phs $M_B$} & \colhead{\phs \csb} &
\colhead{\phs $\log(h)$} & \colhead{\phs $B-V$} & \colhead{\phs ${\cal T}$} }
\startdata
$\log\left(\frac{M_{\rm HI}}{L_B}\right)$ & \phs 0.86 & \phs 0.58
& \phs 0.73 & \phs 0.00 & $-$0.37 & \phs 0.64\nl
$f_g$ & & \phs 0.51 & \phs 0.63 & \phs 0.03 & $-$0.55 & \phs 0.36 \nl
$M_B$ & & & \phs 0.58 & $-$0.59 & $-$0.49 & \phs 0.68 \nl
\csb & & & & \phs 0.18 & $-$0.35 & \phs 0.59 \nl
$\log(h)$ & & & & & \phs 0.33 & $-$0.37 \nl
$B-V$ & & & & & & $-$0.45 \nl
\enddata
\end{deluxetable}

\begin{deluxetable}{lrrr}
\tablewidth{0pt}
\tablecaption{Star Formation Histories \label{tab3}}
\tablehead{
& & \colhead{Exponential\tablenotemark{a}}
& \colhead{Power Law\tablenotemark{b}} }
\startdata
Star Formation Rate\tablenotemark{c,d} & $\dot f_g =$ & $- \tau^{-1}
e^{-t/\tau}$ & $- {{x+1} \over \tau_g} \left({t \over \tau_g} \right)^x$ \nl
Gas Mass Fraction & $f_g =$ & $e^{-t/\tau}$ &
$1-\left({t \over \tau_g} \right)^{x+1}$ \nl
Galaxy Age\tablenotemark{e} & $T_G =$ & $- \tau \ln f_g$
& $\tau_g (1-f_g)^{1/(x+1)}$ \nl
Mean Stellar Age & $\langle T_* \rangle =$ &
$- \tau \left(1+{{\ln f_g} \over {1-f_g}}\right)$ & $T_G/(x+2)$ \nl
Time Scale Limit\tablenotemark{c,f} &
& $\tau > - T_U/\ln f_g$ & \phs $\tau_g < T_U (1-f_g)^{-1/(x+1)}$ \nl
Mean Stellar Age Limit\tablenotemark{f} & $\langle T_* \rangle <$ &
$T_U \left({1 \over {\ln f_g}} + {1 \over {1-f_g}}\right)$
& $T_U/(x+2)$ \nl
\enddata
\tablenotetext{a}{Declining SFR with $\tau > 0$.}
\tablenotetext{b}{Flat or rising SFR with $\tau_g$ and $x \ge 0$.}
\tablenotetext{c}{$\tau$ = $e$-folding time scale; $\tau_g$ = gas consumption
time scale.}
\tablenotetext{d}{Global SFR normalized by total baryonic mass.}
\tablenotetext{e}{$T_G$ defined by the onset of star formation = age of
oldest stars.}
\tablenotetext{f}{Limits obtained by requiring $T_G < T_U$, the age of the
universe.}
\end{deluxetable}

\begin{deluxetable}{lccl}
\tablewidth{0pt}
\tablecaption{Population Models \label{tab4}}
\tablehead{
\colhead{${\cal T}$} & \colhead{$\tau$ (Gyr)} & \colhead{$x$} &
\colhead{Form of SFR} }
\startdata
Sa & \phn 3.3 & \nodata & Exponential decay \nl
Sb & \phn 5 \phd \phn & \nodata & Exponetial decay \nl
Sc & 10 \phd \phn & \nodata & Exponetial decay \nl
Sd & 20 \phd \phn & 0 & Constant \nl
Im & 20 \phd \phn & 2 & Rising power law \nl
\enddata
\tablecomments{Hubble type is used here to denote
the models of Guideroni \& Rocca-Volmerange (1987).
Strictly speaking, these only specify the shape of
the model star formation history.}
\end{deluxetable}

\vspace{4.5in}

\figcaption[mhiB.ps,mhiI.ps]{The HI mass to light ratio as a function of disk
central surface brightness.  a) $B$-band b) $I$-band.
There is a strong correlation between the
gas content of spiral galaxies and their surface brightness.
This is apparent in both passbands even though there are fewer data in $I$.
What scatter there is reflects real differences between galaxies
as the uncertainties are small compared to the
range plotted:  typically 0.2 - 0.3 mag. in \csb, and 0.2 dex in the
distance independent flux ratio $M_{\rm HI}/L_B$.
Note the sharp end to the data as $\csb \rightarrow 25~B\sb$ imposed by
isophotal selection limits.  Without our data for low surface brightness
galaxies, this occurs brighter than $\csb \approx 23$ in (a).  This is
also apparent in the gap in (b) at $\csb^I \approx 21$.  Examining only
galaxies brighter than this conceals the correlation for lack of dynamic
range.  The very low surface brightness giant Malin 1 [the triangle in (a)]
breaks this isophotal barrier because it was serendipitously discovered,
in part due to its high gas content.
\label{mhi}}

\figcaption[MSB.ps,MH.ps]{The optical ($B$-band) properties of spiral
galaxies: a) central surface brightness \vs absolute magnitude and b)
the scale length length \vs absolute magnitude.
Different symbols represent different Hubble types,
progressing from early to late types in order of increasing number of
points on the symbol [key inset in (a)].
\label{MSBH}}

\figcaption[SBVI.ps]{$I$-band central surface brightness
$\csb^I = \csb - (B-I)$ \vs $V-I$
color.  Symbols correspond to Hubble type as per Fig.~2.  Note that
lower surface brightness disks are systematically bluer than those of
higher surface brightness --- they are not the result of fading after
cessation of star formation.
\label{SBVI}}

\figcaption[Bott.ps]{The vertical velocity dispersion of spiral disks as
a function of color.  Data are from Bottema (1993; 1995).  The line is
our estimator for $\Upsilon_*$ mapped onto the velocity dispersion by
equation (8) of Bottema (1993).   The normalization is chosen to
match the data, but the shape of the line is not a fit thereto
(see de Blok \& McGaugh 1997).  Nevertheless, it does match the observed
trend, which goes in the expected sense that redder galaxies have higher
$\Upsilon_*$.
\label{Bott}}

\figcaption[Mstar.ps]{The stellar mass of sample galaxies (in solar masses)
computed from ${\cal M}_* = \Upsilon_* L$ independently in the $B$ (abscissa)
and $I$ (ordinate) bands.  Separate measures of the luminosity ($L_B$ and $L_I$)
are used together with the estimators for $\Upsilon_*$ discussed in the text.
The data closely follow the line of unity, indicating reasonable agreement
between the two methods.
\label{Mstar}}

\figcaption[molec.ps]{The ratio of molecular to atomic gas mass as a function
of Hubble type.  Data (mean and error in the mean) are from Young \& Knezek
(1989).  As usual, the error in the mean understates the amount of real
scatter at a given type.  The line is our fit to the data (see text).
\label{molec}}

\figcaption[fgM.ps,fgS.ps,fgH.ps,fgC.ps]{The gas fraction of
spiral galaxies plotted against a) absolute $B$ magnitude,
b) disk central surface brightness, c) disk scale length and d) color.
The full possible range of $0 \leq f_g \leq 1$ is plotted and
for the first time is needed --- the dimmest
galaxies are mostly gaseous with $f_g \gtrsim 0.5$.  A strong
relation exists between $f_g$ and $M_B$, \csb, and color.
There is no relation at all between $f_g$ and the size of the disk.
Note that the $f_g$-$M_B$ and -\csb\ relations have the same slope.
Malin 1 (triangle) is an extreme case with a gas fraction which
is abnormally high for its luminosity but normal for its surface brightness.
\label{fg}}

\figcaption[fgMI.ps,fgSI.ps,fgHI.ps,fgCI.ps]{Like Fig.~7
except for the $I$-band data.  The figure shows $f_g^I$ against
a) absolute $I$ magnitude, b) $I$-band disk central surface brightness,
c) scale length, and d) $V-I$ color.
The gas fraction is determined independently from that in Fig.~7
by assuming that $\Upsilon_*^I$ is constant for all spirals.
This gives indistinguishable
results from the $B$-band even though a completely different estimator
for the stellar mass has been employed.
\label{fgI}}

\figcaption[Tracks.ps]{The data from
Fig.~7 (a) and (b) overplotted after shifting
by the intercepts of equation~(3).  Circles: $M_B$.  Squares: \csb.
Lines are model evolutionary tracks from Guideroni \& Rocca-Volmerange (1987).
Horizontal lines are isochrones every 2 Gyr from 1 to 15 Gyr.
Vertical lines are evolutionary trajectories for specific star
formation histories, labeled by the Hubble type the model attempts to
represent.   Early types (Sa to Sc) have exponentially declining
star formation rates, while Sd is constant and Im rising with time
(Table~4).  The position of these lines depends on the total baryonic mass
of a galaxy.  They are drawn with separations corresponding to the mean
of each type as stipulated by Guideroni \& Rocca-Volmerange (1987); the
absolute position of the resulting grid has been adjusted to coincide
with the data.  Though these (or any) population synthesis models are by
no means perfect, they do illustrate how $f_g$ constrains
the combination of evolutionary rate and age.  With a large dynamic
range in $f_g$ and the traditional constraints of color and metallicity,
it may be possible to construct a multidimensional extragalactic analog
to the HR diagram which might enable the age-dating of spiral galaxies.
\label{Tracks}}


\clearpage

\begin{references}

\reference{HDF} Abraham, R. G., Tanvir, N. R., Santiago, B. X.,
	Ellis, R. S., Glazebrook, K., \& van den Bergh, S. 1996,
	\mnras, 279, 47

\reference{BF} Babul, A., \& Ferguson, H. C. 1996, \apj, 458, 100

\reference{BR} Bergvall, N., \& R\"onnback, J. 1995, \mnras, 273, 603

\reference{BC} Bruzual, A. G., \& Charlot, S. 1993, \apj, 405, 538

\reference{B85} Bothun, G.~D. 1985, \aj, 90, 1982

\reference{malin1} Bothun, G.~D., Impey, C.~D., Malin, D.~F., \& Mould, J.~R.
        1987, \aj, 94, 23

\reference{B93} Bothun, G.~D., Schombert, J.~M., Impey, C.~D.,
      Sprayberry, D., \& McGaugh, S.~S. 1993, \aj, 106, 530

\reference{Bott1} Bottema, R. 1993, \aap, 275, 16

\reference{Bott2} Bottema, R. 1995, Ph.D. thesis, University of Groningen

\reference{D93} Dalcanton, J. J. 1993, \apj, 415, L87

\reference{chain} Dalcanton, J. J., \& Shectman, S. A. 1996, \apj, 465, L9

\reference{dB1} de~Blok, W.~J.~G., van~der~Hulst, J.~M.,
	\& Bothun, G.~D. 1995, \mnras, 274, 235

\reference{dBM} de~Blok, W.~J.~G., \& McGaugh, S.~S. 1996, \apj, 469, L89

\reference{dBM2} de~Blok, W.~J.~G., \& McGaugh, S.~S. 1997, \mnras, submitted

\reference{dB2} de~Blok, W.~J.~G., McGaugh, S.~S., \& van~der~Hulst, J.~M.
	1996, \mnras, in press

\reference{dJt} de~Jong, R.~S. 1995, Ph.D. thesis, University of Groningen

\reference{dJ3} de~Jong, R.~S. 1996, \aap, 313, 45

\reference{dJ1} de~Jong, R.~S., \& van der Kruit, P.~C. 1994, A\&AS, 106, 451

\reference{DP} Disney, M.~J., \& Phillipps, S. 1983, \mnras, 205, 1253

\reference{driver} Driver, S., Windhorst, R. A., \& Griffiths, R. E. 1995
	453, 48

\reference{ECBHG} Ellis, R. S., Colless, M., Broadhurst, T. J., Heyl, J. S.,
	\& Glazebrook, K. 1996, \mnras, 280, 235

\reference{harry} Ferguson, H.~C., \& McGaugh, S.~S. 1995, \apj, 440, 470

\reference{IM} Im, M., Ratnatunga, K. U., Griffiths, R. E., \& Casertano, S.
	1995, \apj, 445, L15

\reference{IB} Impey, S. D., \& Bothun, C. D. 1989, \apj, 341, 89

\reference{imp} Impey, S. D., Sprayberry, D., Irwin, M., \& Bothun, C. D. 1996,
	preprint

\reference{GRV} Guideroni \& Rocca-Volmerange 1987, \aap, 186, 1

\reference{kgb} Glazebrook, K., Ellis, R., Santiago, B., \& Griffiths, R.
	1995, \mnras, 275, L19

\reference{jeremy} Heyl, J., Colless, M., Ellis, R. S., \&
	Broadhurst, T. 1996, \mnras, in press

\reference{K89} Kennicutt, R.~C. 1989, \apj, 344, 685

\reference{KTC} Kennicutt, R. C., Tamblyn, P., \& Congdon, C. E. 1994,
	\apj, 435, 22

\reference{CFRS} Lilly, S. J., Tresse, L., Hammer, F., Crampton, D.,
	\& Le F\`evre, O. 1995, \apj, 455, 108

\reference{mythesis} McGaugh, S.~S. 1992, Ph.D. thesis, University of Michigan

\reference{OHme} McGaugh, S.~S. 1994a, \apj, 426, 135

\reference{FBGs} McGaugh, S.~S. 1994b, \nat, 367, 538

\reference{me} McGaugh, S.~S. 1996a, \mnras, 280, 337

\reference{IAU} McGaugh, S.~S. 1996b, in IAU Symposium 171: New Light
	on Galaxy Evolution, eds. R. Bender \& R. L. Davies,
	(Dordrecht:  Kluwer), 97

\reference{MB} McGaugh, S.~S., \& Bothun, G.~D. 1994, \aj, 107, 530

\reference{MBS} McGaugh, S.~S., Bothun, G.~D., \& Schombert, J.~M. 1995a,
	\aj, 110, 573

\reference{LSBmorph} McGaugh, S.~S., Schombert, J.~M., \& Bothun, G.~D.
	1995b, \aj, 109, 2019

\reference{MLR} McLeod, B. A., \& Rieke, M. J. 1995, \apj, 454, 611

\reference{mo} Mo, H.~J., McGaugh, S.~S., \& Bothun, G.~D. 1994, \mnras,
	267, 129

\reference{nnet} Naim, A., Lahav, O., Sodr\'e, L., Storrie-Lombardi, M. C.
	1995, \mnras, 275, 567

\reference{PD} Phillipps, S., \& Driver, S. 1995, \mnras, 274, 832

\reference{Pozz} Pozzetti, L., Bruzual, G., \& Zamorani, G. 1996,
	\mnras, in press

\reference{RB} Rao, S. \& Briggs, F. 1993, \apj, 419, 515

\reference{RX} Roberts, M.~S. 1963, \araa, 1, 149

\reference{rom2} Romanishin, W., Krumm, N., Salpeter, E. E., Knapp, G. R.,
	Strom, K. M., \& Strom, S. E. 1982, \apj, 263, 94

\reference{rom1} Romanishin, W., Strom, K. M., \& Strom, S. E. 1983,
	\apjs, 53, 105

\reference{RB1} R\"onnback, J., \& Bergvall, N. 1994, A\&AS, 108, 193

\reference{RB2} R\"onnback, J., \& Bergvall, N. 1995, \aap, 302, 353

\reference{SH} Salpeter, E. E., \& Hoffman, G. L. 1996, 465, 595

\reference{CNOC} Schade, D., Carlberg, R. G., Yee, H. K. C., Lopez-Cruz,
	O., \& Ellingson, E. 1996, \apj, 465, 103

\reference{schade} Schade, D., Lilly, S. J., Crampton, D., Hammer, F.,
	Le F\`evre, O., \& Tresse, L. 1995, \apj, 451, L1

\reference{CO} Schombert, J.~M., Bothun, G.~D., Impey, C.~D., \& Mundy,
	L.~G. 1990, \aj, 100, 1523

\reference{lsbcat} Schombert, J.~M., Bothun, G.~D., Schneider, S.~E., \& 
        McGaugh, S.~S. 1992, \aj, 103, 1107

\reference{T68} Tinsley, B. M. 1968, \apj, 151, 547

\reference{vdH} van der Hulst, J. M., Skillman, E. D., Smith, T. R.,
	Bothun, G. D., McGaugh, S. S., \& de Blok, W. J. G. 1993, \aj, 106, 548

\reference{Wils} Wilson, C. D. 1995, \apj, 448, L97

\reference{Guy} Worthey, G. 1994, \apjs, 95, 107

\reference{H2H1} Young, J. S., \& Knezek, P. M. 1989, \apj, 347, L55

\end{references}
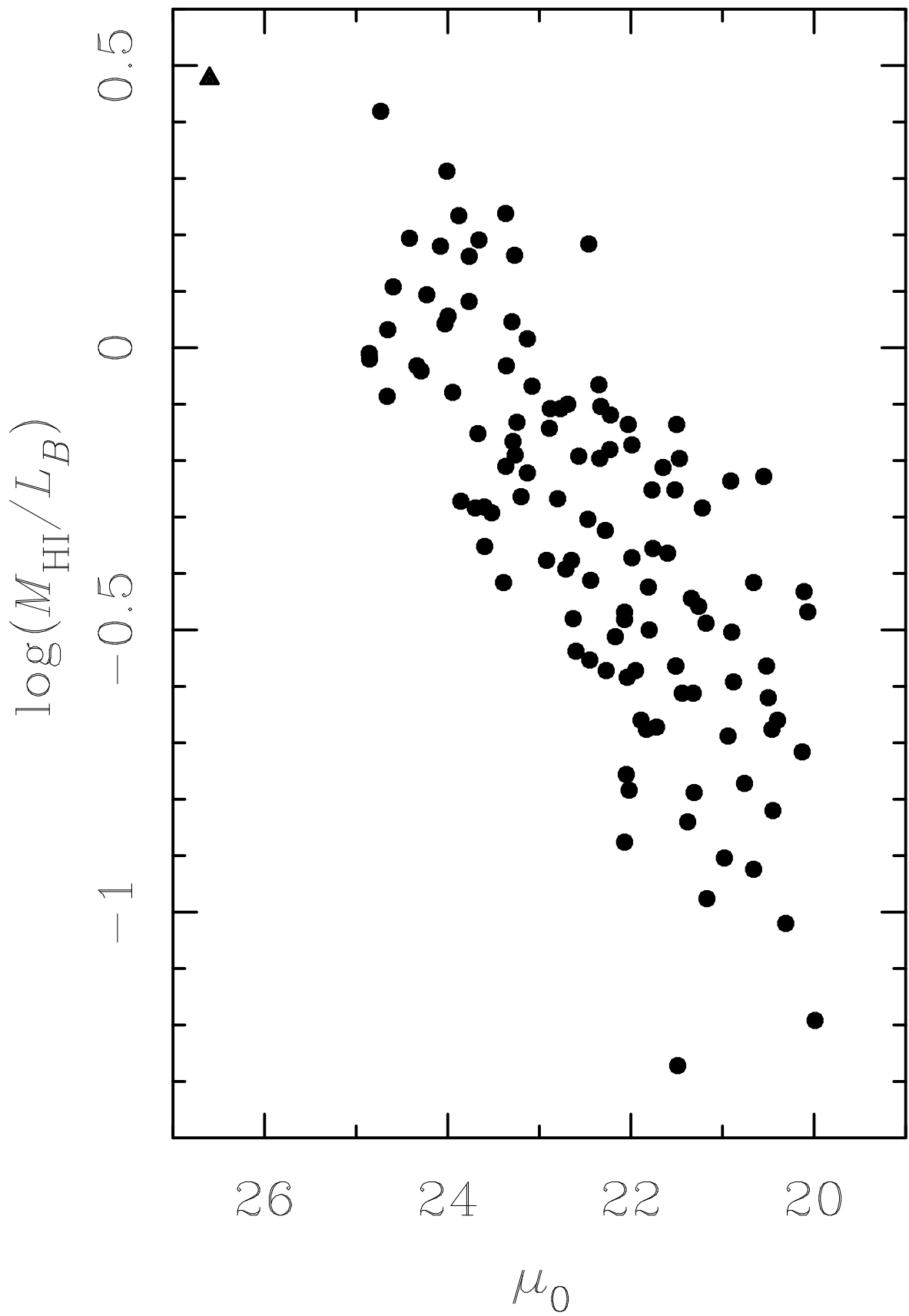
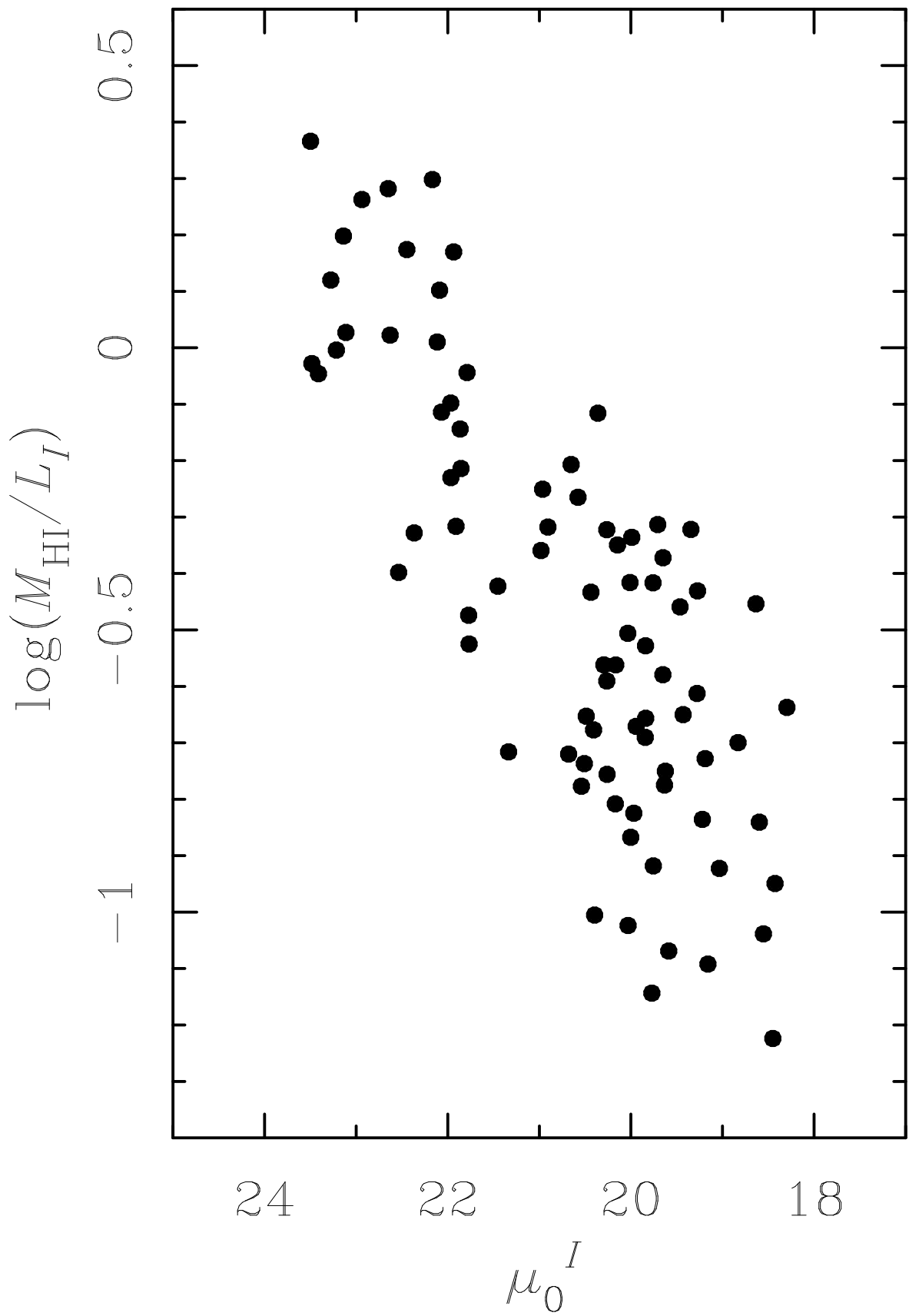
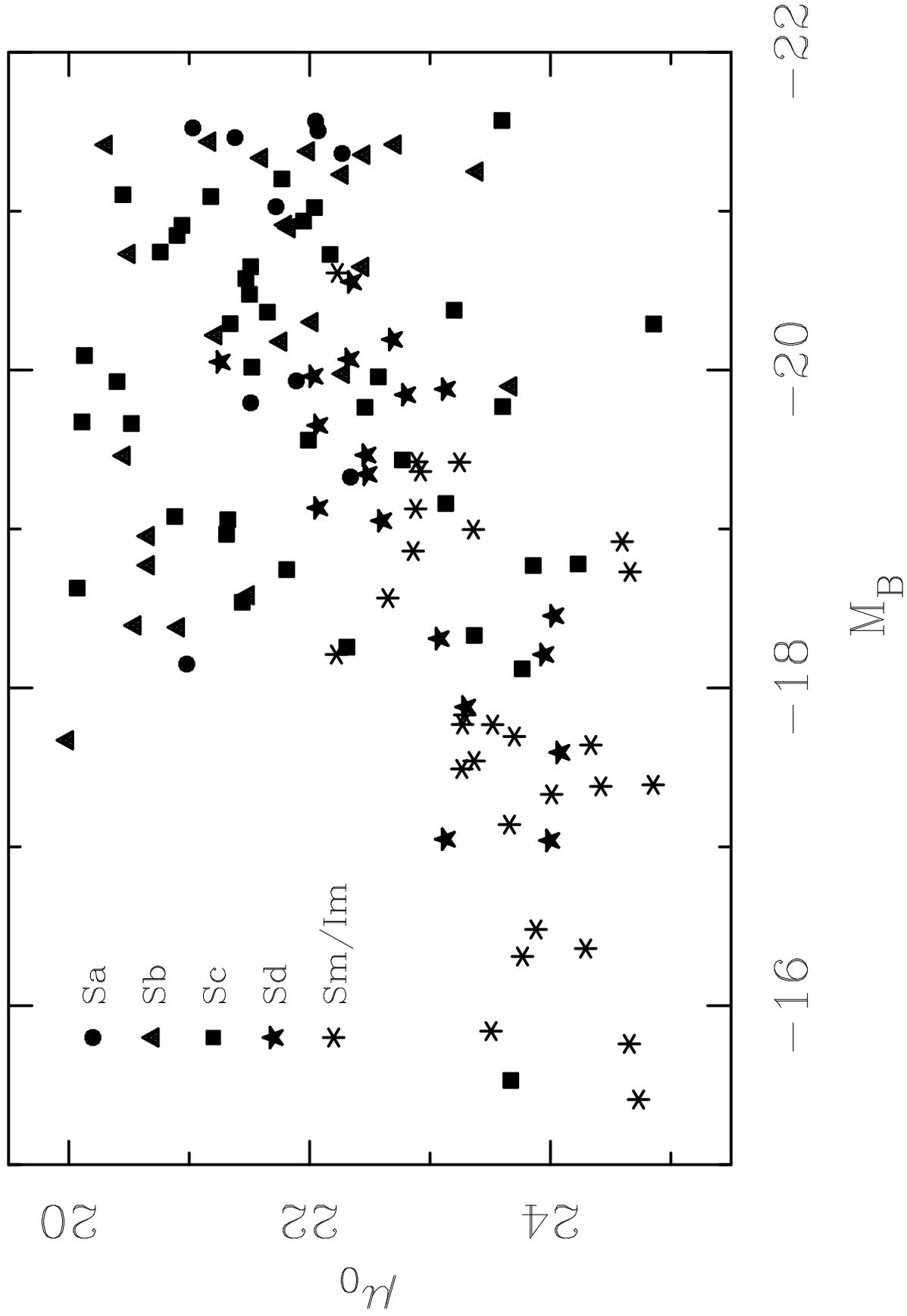
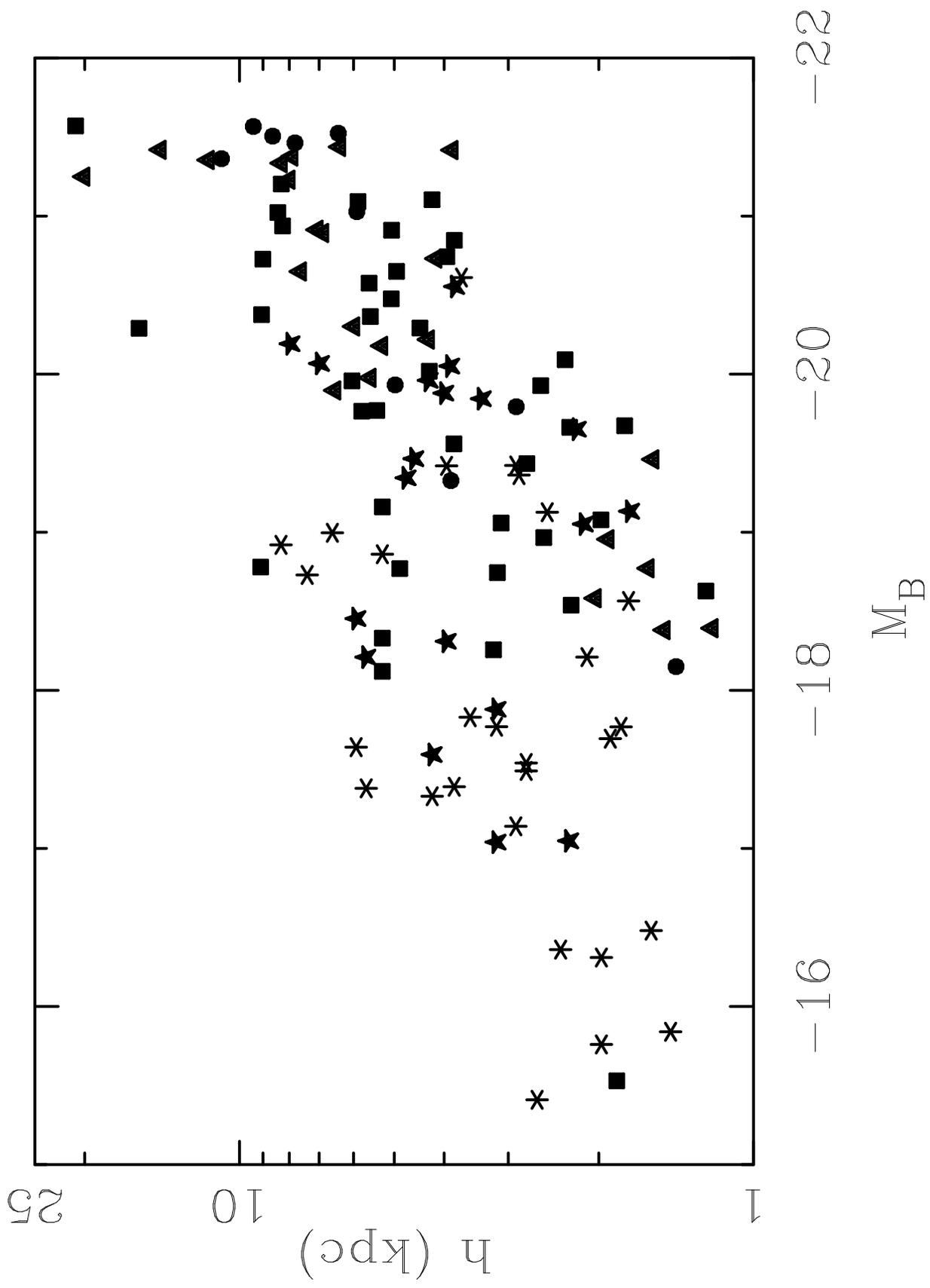
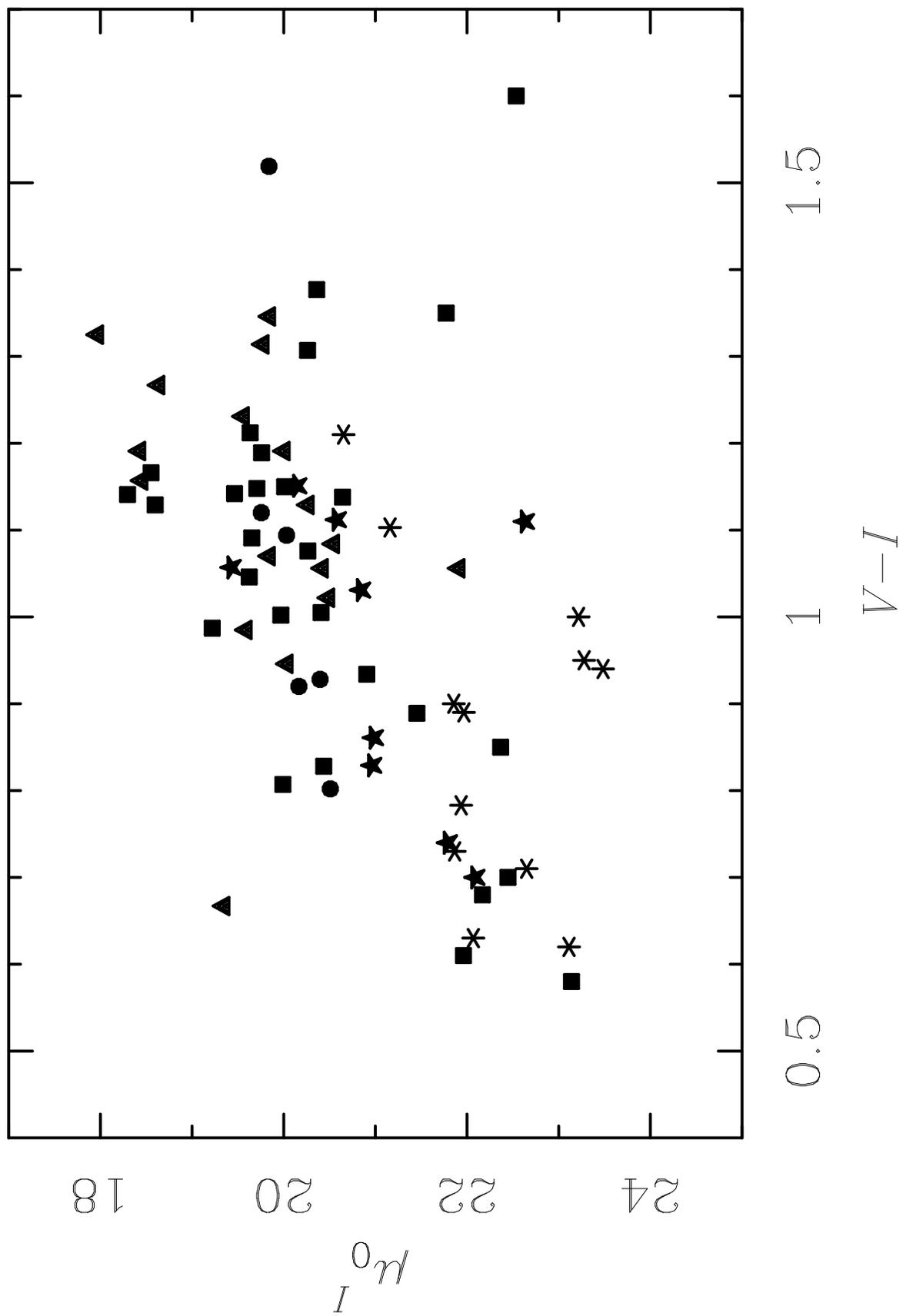
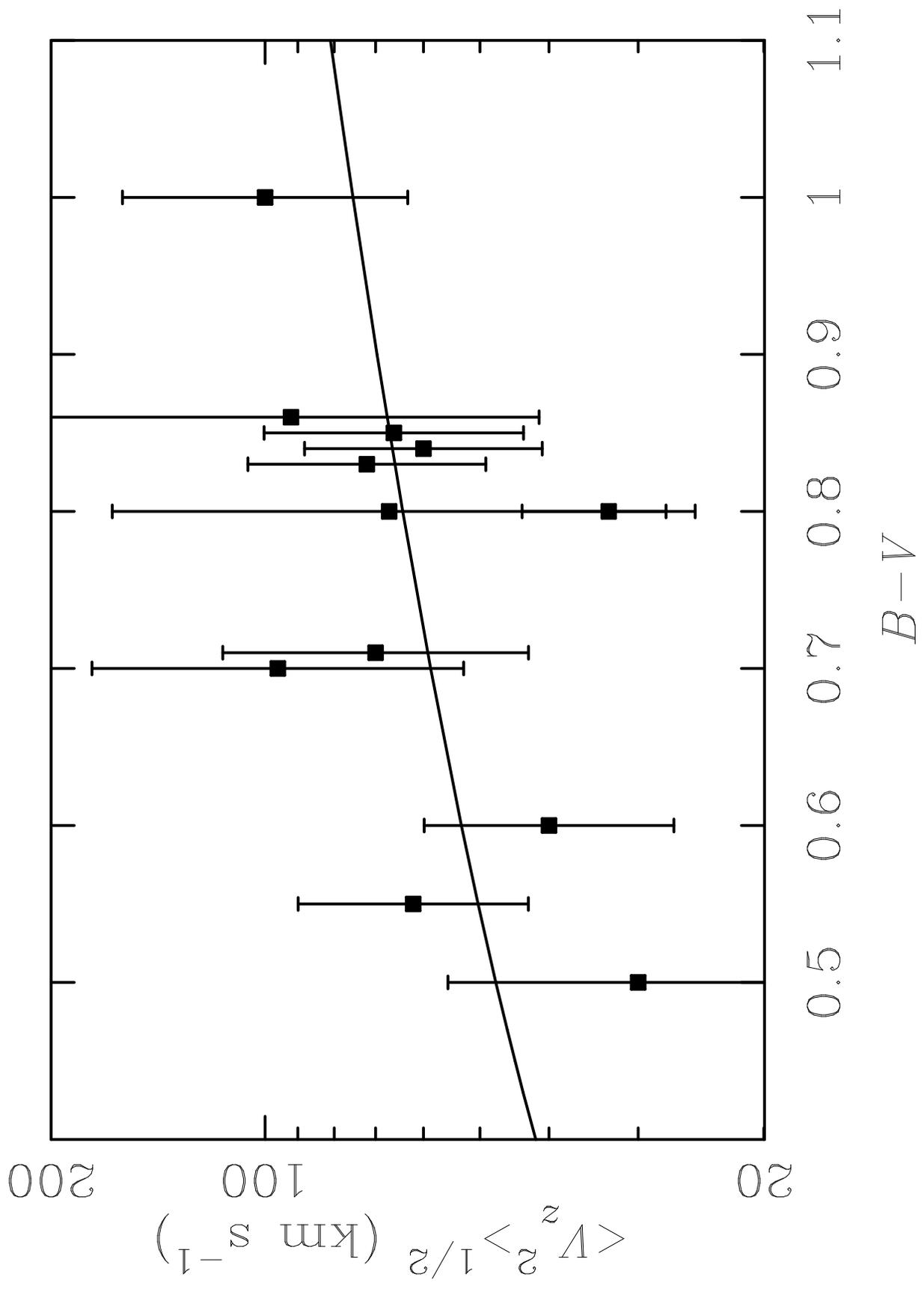
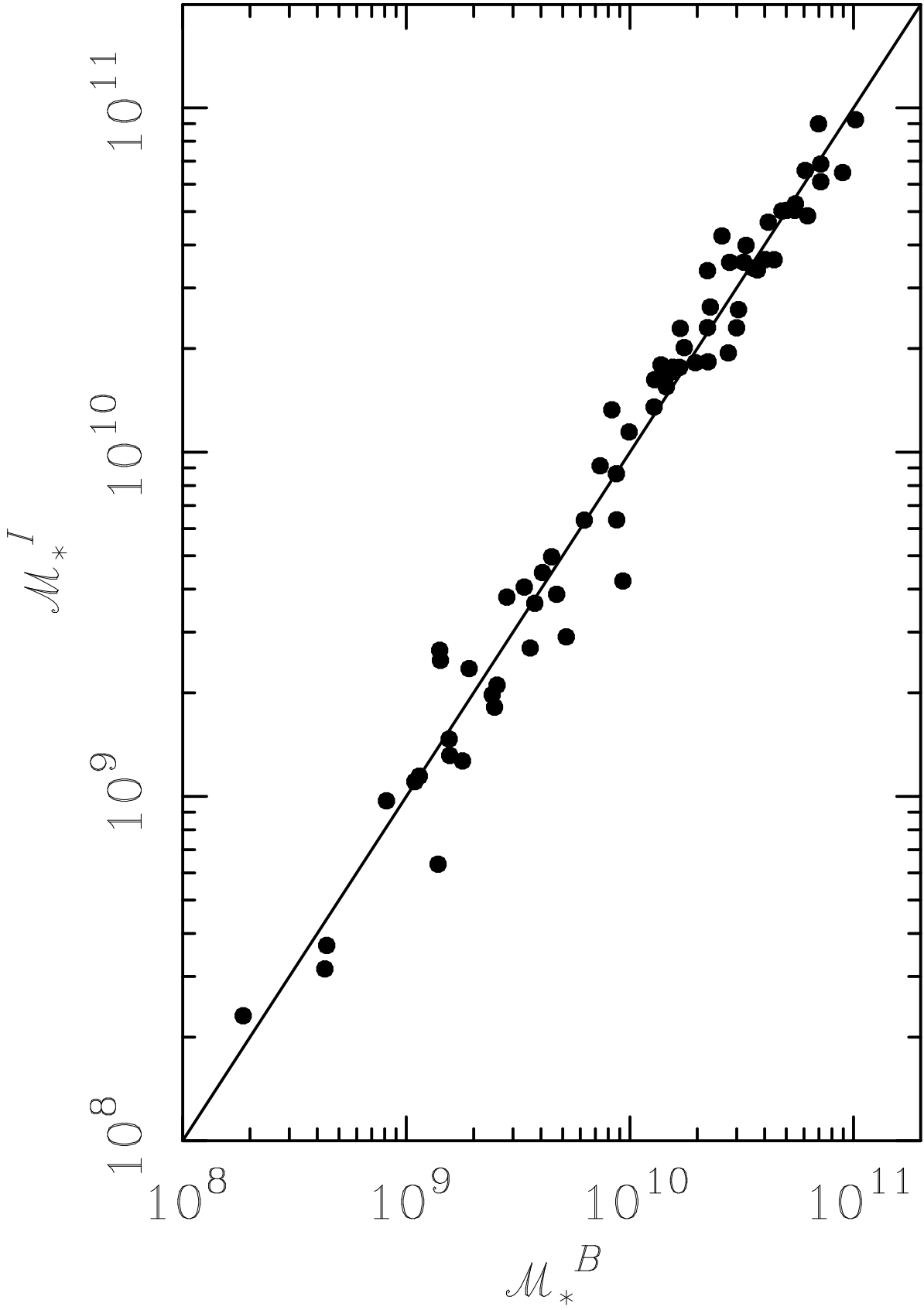
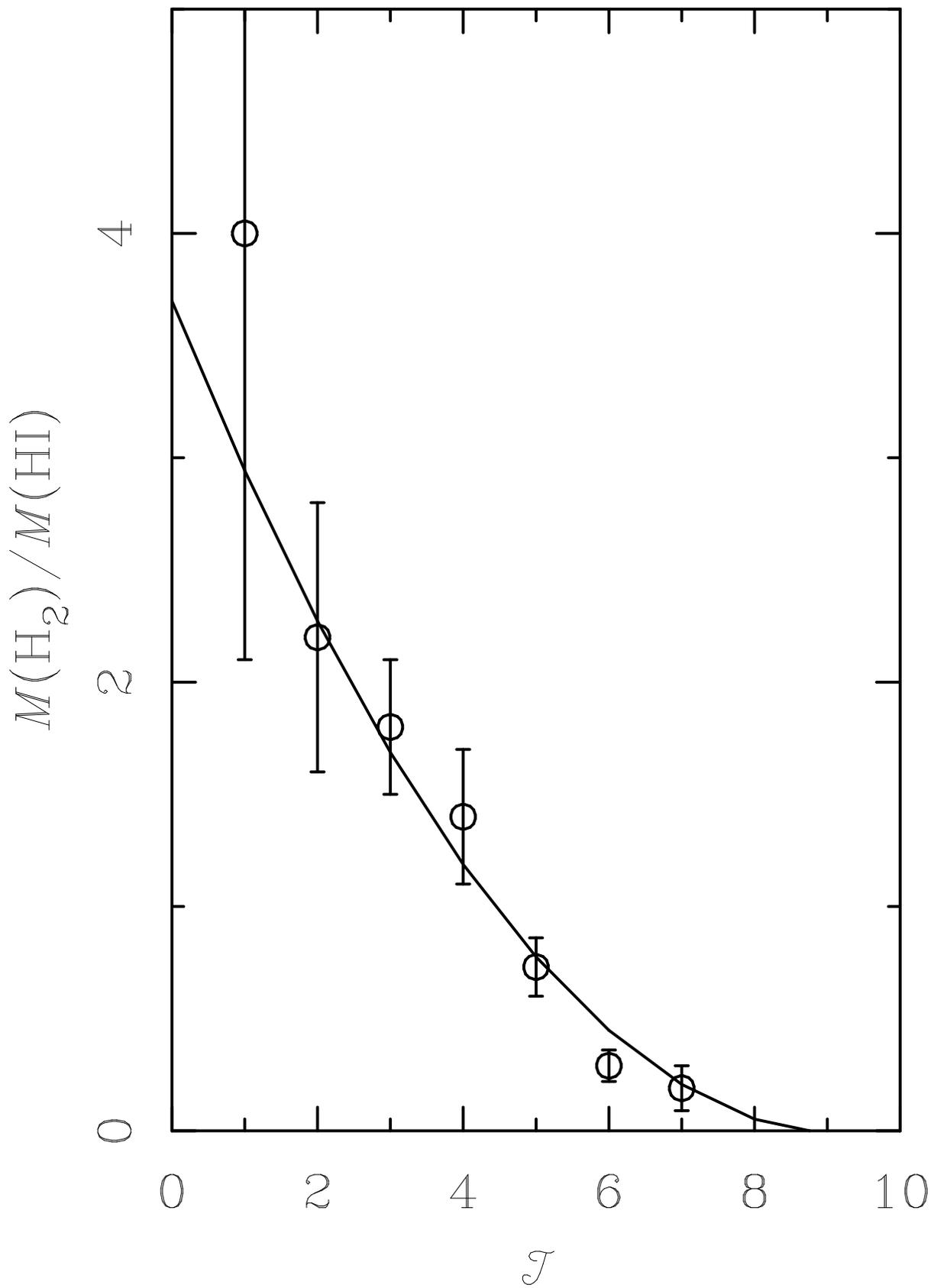
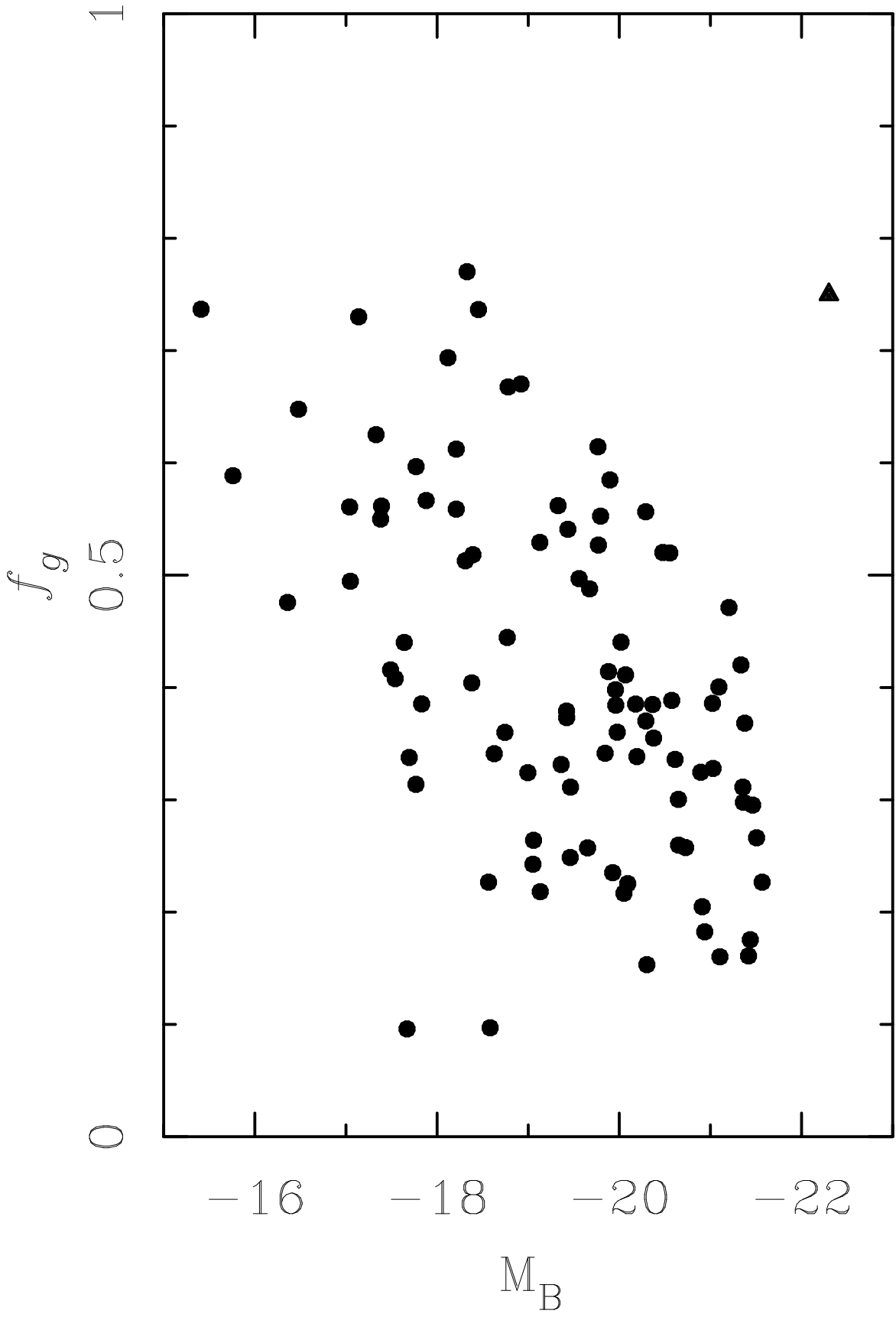
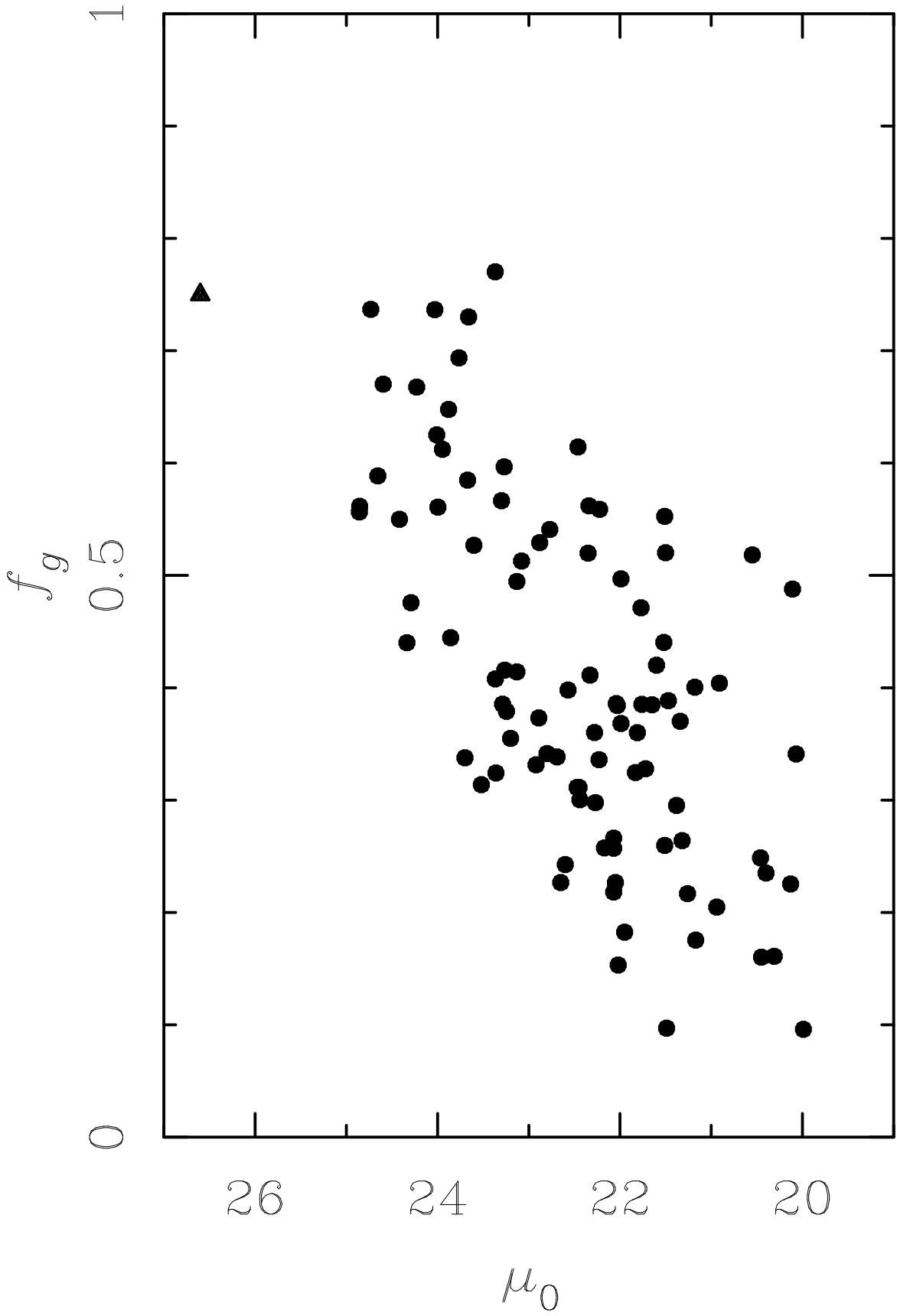
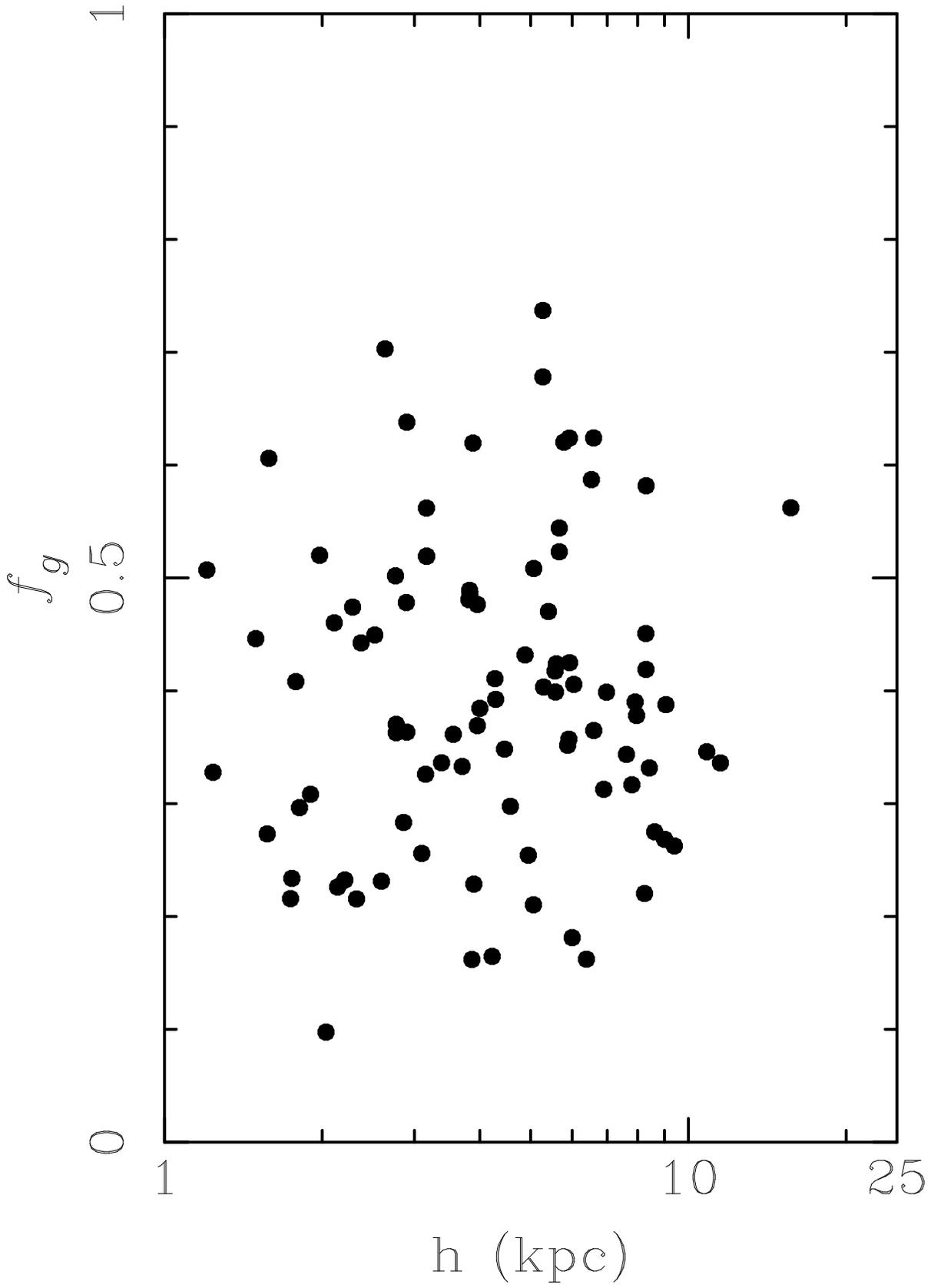
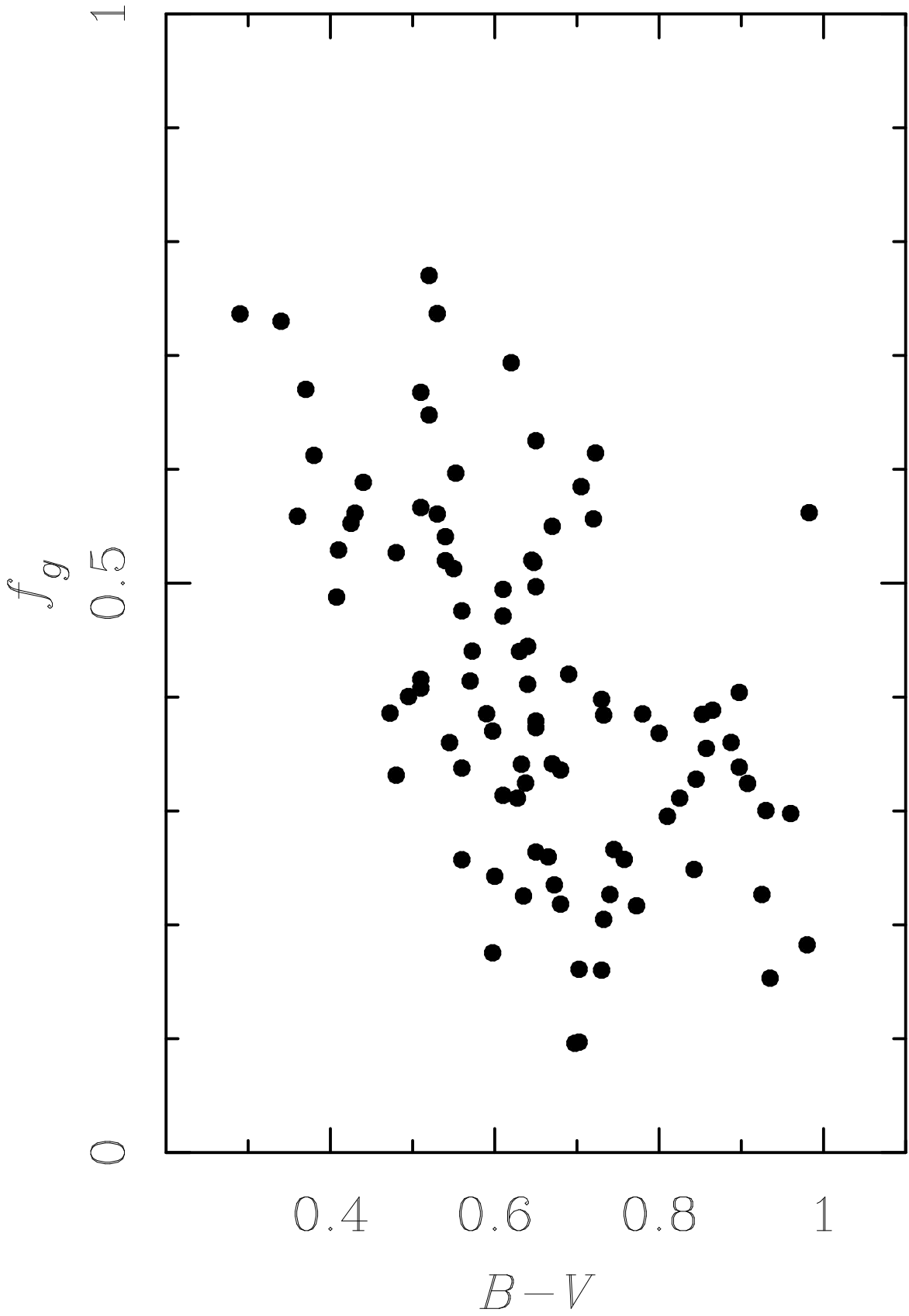
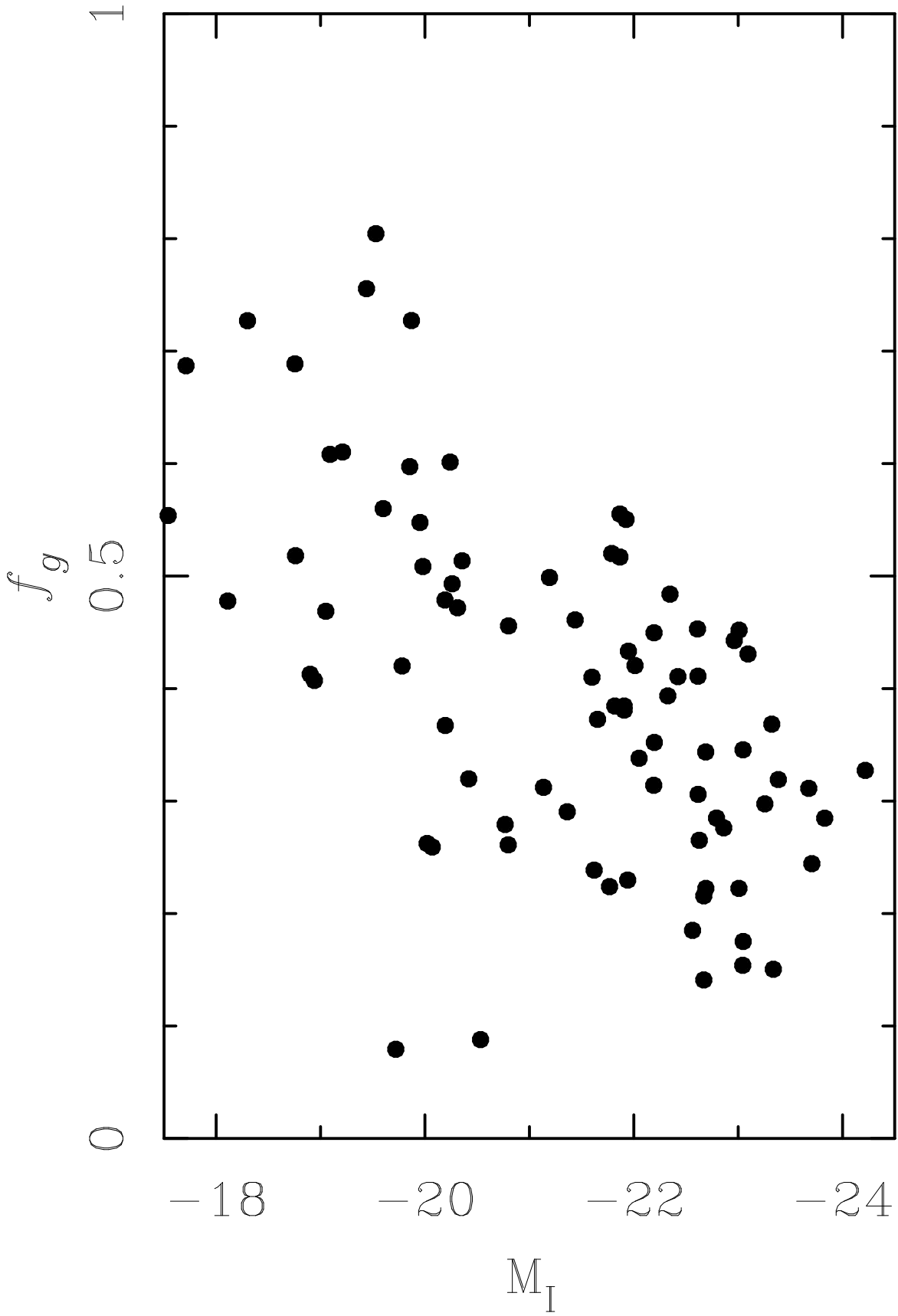
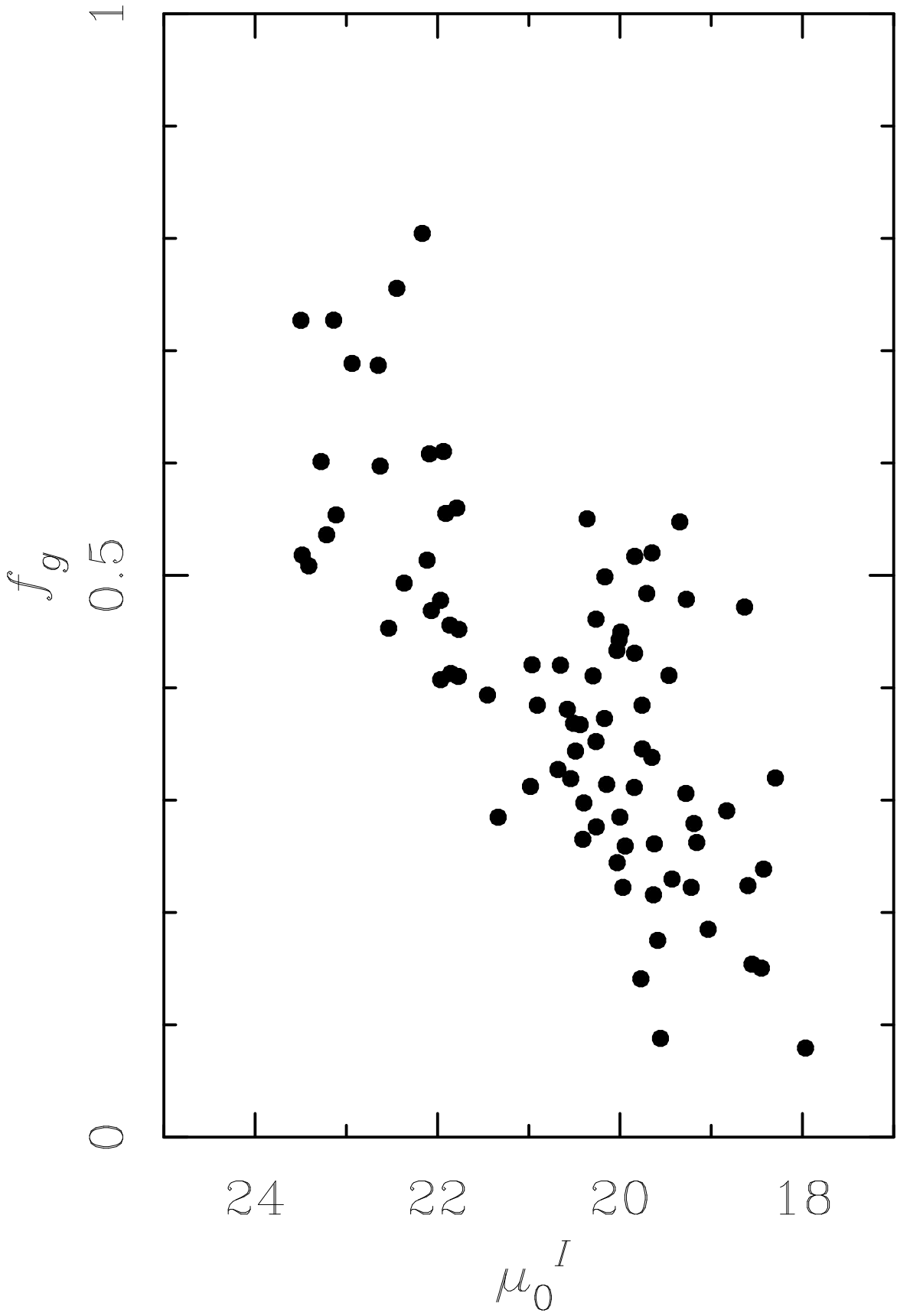
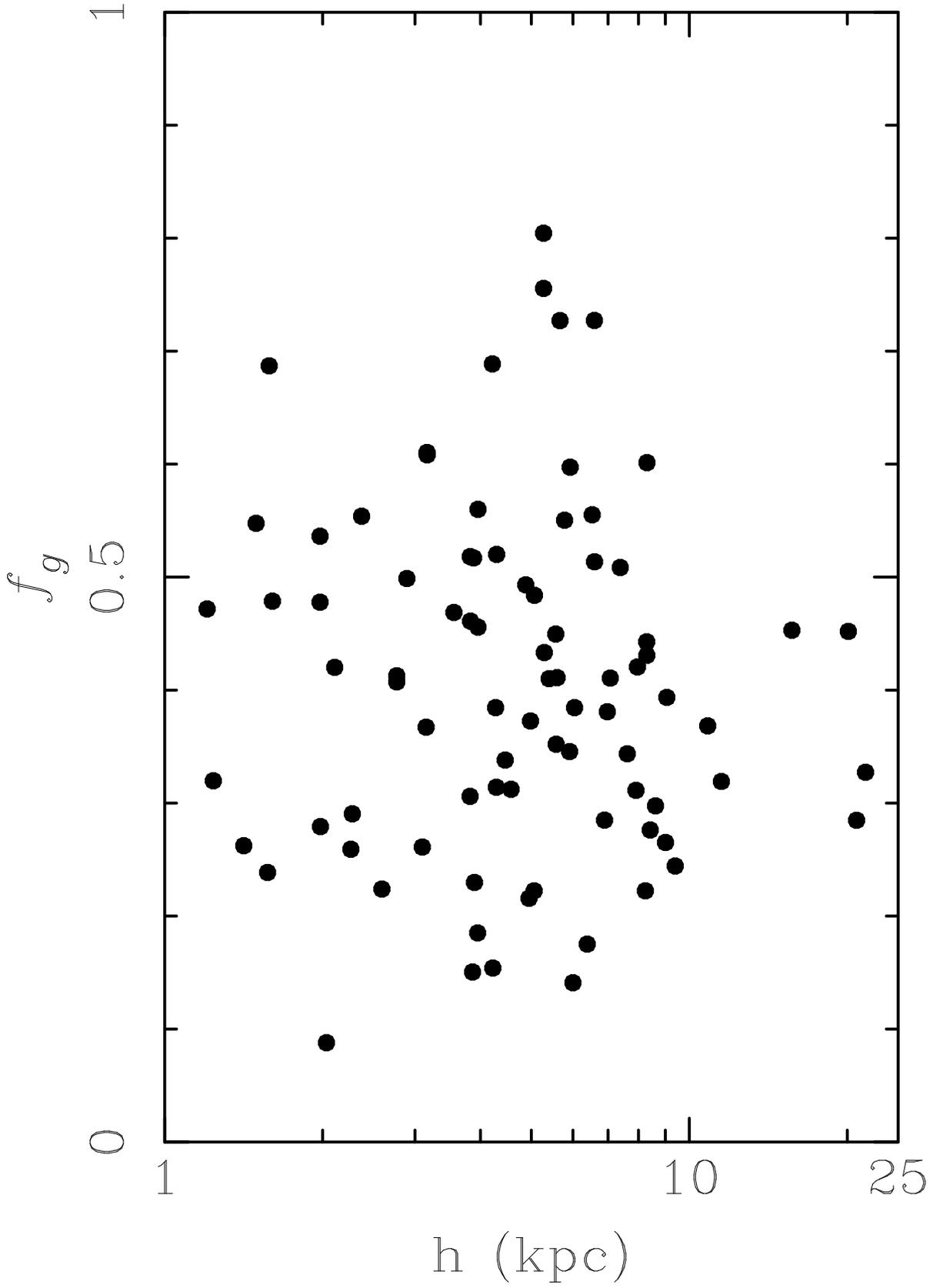
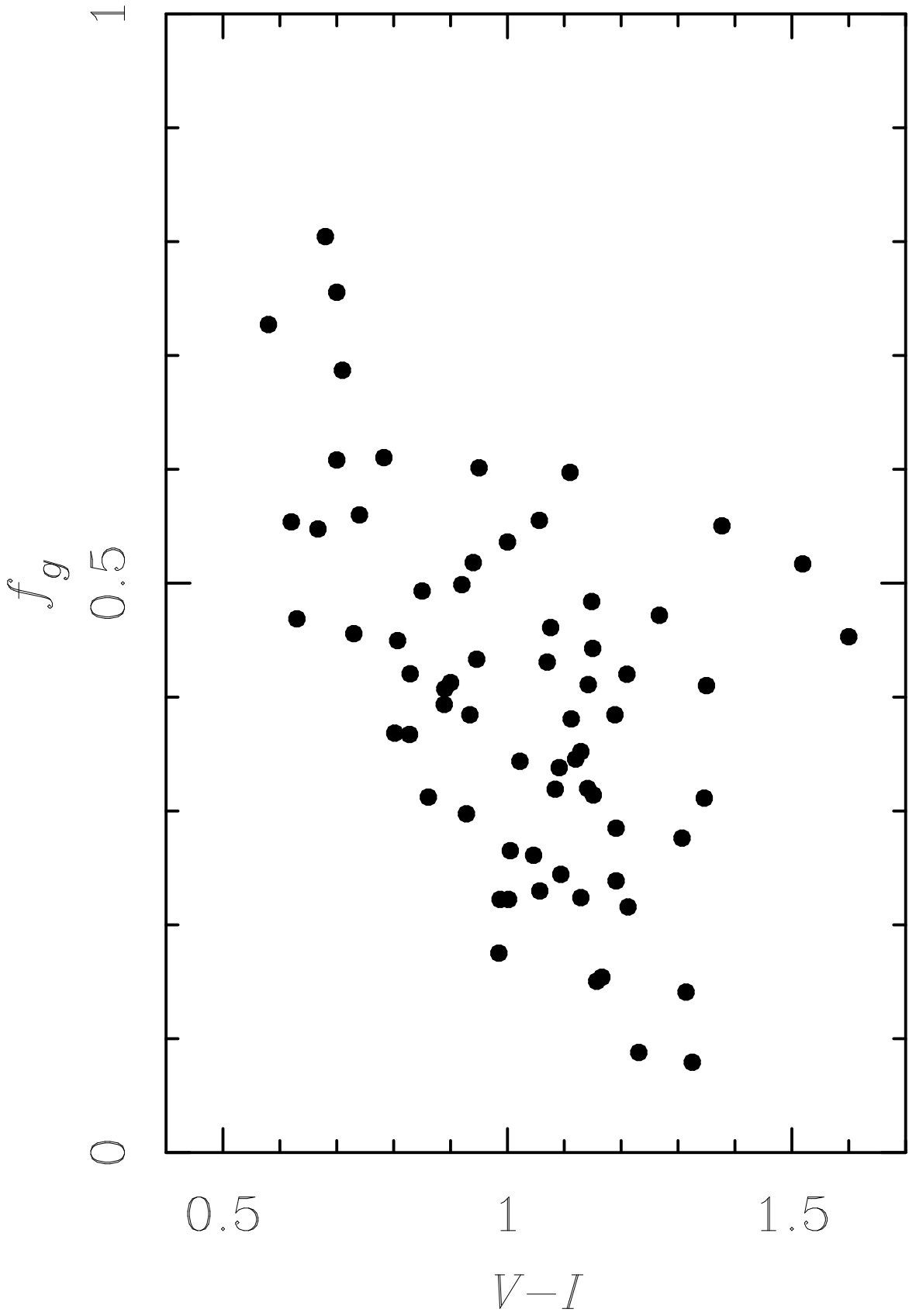
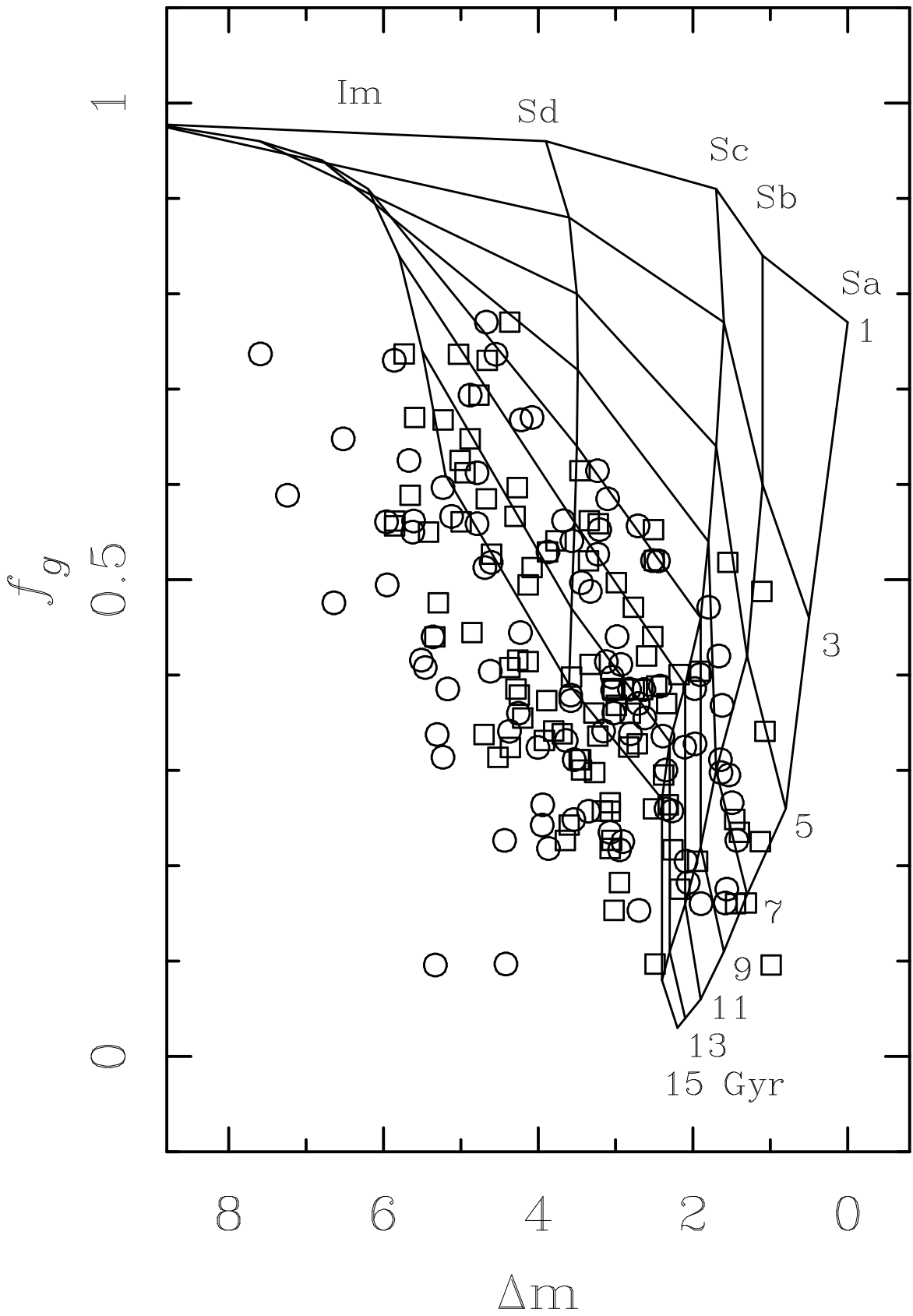
\end{document}